\documentclass[external]{esp-paper}
\usepackage{amsmath}

\usepackage[utf8]{inputenc} %
\usepackage[T1]{fontenc}
\usepackage{placeins}%
\usepackage{hyperref}       %
\usepackage{url}            %
\usepackage{booktabs}       %
\usepackage{amsfonts}       %
\usepackage{nicefrac}       %
\usepackage[table]{xcolor}  %
\usepackage{longtable}
\usepackage{multirow}
\usepackage{booktabs}
\usepackage{graphicx}
\usepackage{tabularx}

\usepackage[most]{tcolorbox}
\usepackage{enumitem}
\usepackage{subcaption}
\usepackage{multicol}
\usepackage{enumitem}

\newcommand{\headinggap}{%
  \par\addvspace{0.6\baselineskip}%
  \Needspace{3\baselineskip}%
}
\titlespacing*{\paragraph}{0pt}{0.6\baselineskip}{1em}

\definecolor{PineGreen}{HTML}{0B3D2E}
\definecolor{SageGreen}{HTML}{C8D5B9}
\definecolor{LightGray}{HTML}{F4F4F2}
\hypersetup{
  colorlinks=true,
  linkcolor=PineGreen,
  citecolor=PineGreen,
  urlcolor=PineGreen,
  pdftitle={BEANS-Next and ROOTS: Broadening Audio-Language Capabilities for Bioacoustics},
  pdfauthor={Christos Plachouras; David Robinson; Marius Miron; Gagan Narula; Paul Laisné; Anthony L. T. Fine; Benno Weck; Ellen Gilsenan-McMahon; Diane Kim; Laura Hay Mack; Maddie Cusimano; Sara Keen; Lukas Rauch; Benjamin Hoffman; Emmanuel Chemla; Emmanouil Benetos; Johan Pauwels; Milad Alizadeh; Matthieu Geist; Olivier Pietquin}
}

\newcommand{\benchmark}{BEANS-Next}
\newcommand{\dataset}{ROOTS}
\newcommand{\model}{NatureLM-audio v1.1}

\title{BEANS-Next and ROOTS:\\Broadening Audio-Language Capabilities for Bioacoustics}
\shorttitle{BEANS-Next and ROOTS: Broadening Audio-Language Capabilities for Bioacoustics}

\makeatletter
\preto{\@author}{\raggedright}
\makeatother

\author[1,2]{\mbox{Christos Plachouras}}
\author[1]{\mbox{David Robinson}}
\author[1]{\mbox{Marius Miron}}
\author[1]{\mbox{Gagan Narula}}
\author[1]{\mbox{Paul Laisn\'e}}
\author[1]{\mbox{Anthony L. T. Fine}}
\author[1]{\mbox{Benno Weck}}
\author[1]{\mbox{Ellen Gilsenan-McMahon}}
\author[1]{\mbox{Diane Kim}}
\author[1]{\mbox{Laura Hay Mack}}
\author[1]{\mbox{Maddie Cusimano}}
\author[1]{\mbox{Sara Keen}}
\author[1]{\mbox{Lukas Rauch}}
\author[1]{\mbox{Benjamin Hoffman}}
\author[1]{\mbox{Emmanuel Chemla}}
\author[2]{\mbox{Emmanouil Benetos}}
\author[2]{\mbox{Johan Pauwels}}
\author[1]{\mbox{Milad Alizadeh}}
\author[1]{\mbox{Matthieu Geist}}
\author[1]{\mbox{Olivier Pietquin}}
\affil[1]{Earth Species Project}
\affil[2]{Queen Mary University of London}
\date{}

\begin{document}

\thispagestyle{firstpage}
\maketitle

\begin{abstract}

Bioacoustics and ethology encompass a wide range of audio understanding tasks, many of which stand to benefit from recent advances in large audio-language models. However, progress in the field has so far been assessed on a narrow set of tasks, primarily centered on label-centric biological category recognition, such as species and call-type classification. In this work, we introduce BEANS-Next, a benchmark grounded in a taxonomy of bioacoustics tasks spanning acoustic perception, biological category recognition, scene understanding, and in-context learning. Using BEANS-Next, we show that existing models exhibit limited performance beyond the task families emphasized by existing evaluations, constraining their usefulness for broader bioacoustic applications. To support progress on this broader task space, we also introduce ROOTS, a large-scale training resource built from expanded curated real-world data and previously underused behavioral and acoustic metadata, supplemented by audio-derived information and scalable synthetic generation where labeling is insufficient. We demonstrate that training on this dataset yields substantial progress across all task groups of BEANS-Next, moving audio-language models closer to their potential as general-purpose assistants for bioacoustics and ethology. To accelerate progress in the field, we open-source our benchmark, dataset, and data pipelines.\footnote{Project page: \url{https://earthspecies.github.io/beans-next/}.}

\end{abstract}

\section{Introduction}

Bioacoustics examines how animals produce, perceive, and use sound. Animal sounds carry information about species identity, behavior, social dynamics, and ecological context, making bioacoustics central to animal behavior research, biodiversity monitoring, conservation, and the study of animal communication \citep{bradbury1998principles, fischer2013bioacoustic, rutz2023using}. Machine learning has become a standard tool for analyzing these sounds, with applications including species identification, individual recognition, call-type classification, and behavioral annotation \citep{stowell2022computational}. However, the practical needs of ethologists---researchers who study animal behavior---are broader: researchers often measure acoustic structure \citep{kim2025wow}, identify biological categories, count and compare events \citep{sueur2008rapid, holt2009speaking, briggs2012acoustic, marques2013estimating}, reason over temporal patterns \citep{kershenbaum2024, izard2025}, and adapt analyses to new taxa or study-specific sound classes \citep{brown2021, nieto2023systematic, cauzinille2024primate, kate2025, kyalo2025insect}. This breadth makes bioacoustics a compelling domain for large audio-language models (LALMs), which promise a single interface for diverse audio-understanding tasks.

Current LALMs for bioacoustics lack the breadth required to address the full needs of practitioners. Existing benchmarks primarily evaluate classification into a limited subset of biologically relevant categories, especially species or a narrow set of call types \citep{hagiwara2023beans, rauchbirdset}. Many core bioacoustic abilities remain untested: acoustic measurement, call counting, temporal reasoning, call comparison, and adaptation to new taxa or task-specific categories \citep{sueur2008rapid, holt2009speaking, briggs2012acoustic}. At the same time, standard training datasets provide limited supervision for these abilities. Bioacoustic archives are often weakly labeled and semi-structured, making them difficult to incorporate into audio-language model training without substantial processing into scientifically meaningful audio-language tasks \citep{vellinga2015xeno, perez2025wabad, chasmai2024inaturalist, liang2024mind}. This bottleneck means we currently lack both the evaluation data to measure missing abilities in audio-language models and the training resources needed to make progress.

\looseness=-1
\begin{figure}
    \centering
    \includegraphics[width=1\linewidth]{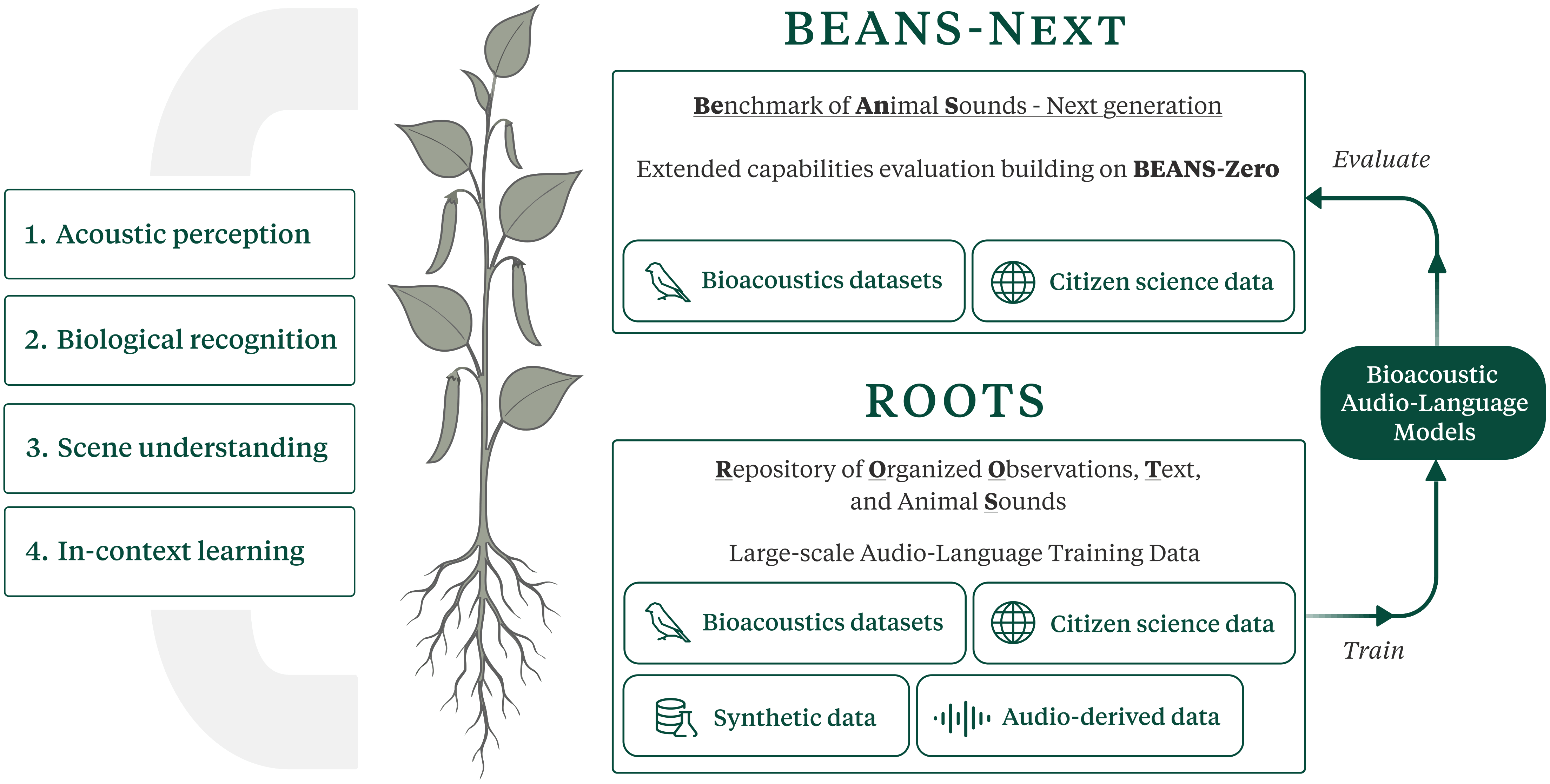}
    \caption{Overview of \textbf{\benchmark{}} and \textbf{\dataset{}}, along with their underlying data sources and the shared four-tier task taxonomy that grounds them.}
    \label{fig:overview}
\end{figure}

To address the evaluation side of this bottleneck, we introduce \benchmark, a benchmark for measuring broad bioacoustic audio-language abilities needed for ethology research. \benchmark{} is organized around a taxonomy of tasks that reflects both scientific workflows and groups of audio-language model capabilities, spanning acoustic perception, biological category recognition, scene understanding, and in-context learning. It builds on established benchmarks BEANS-Zero \citep{robinsonnaturelm} and BirdSet \citep{rauchbirdset}, while broadening them across all task groups. Evaluating on \benchmark{} reveals that existing bioacoustic models are narrow, performing competitively on familiar tasks but poorly on broader ones, while general-purpose audio-language models remain poor across most tasks.

To address the training side of this bottleneck, we introduce \dataset, a large-scale bioacoustic training dataset with 44 million audio-language pairs from 6.5 million distinct audio clips. \dataset{} is built from curated bioacoustic archives, acoustic and behavioral metadata that are underused in computational research, scalable synthetic audio pipelines, and audio-derived metadata. We convert semi-structured and unstructured metadata, such as field notes, into instruction-following tasks for species recognition, taxonomy prediction, call-type classification, life-stage prediction, acoustic description, and other biologically meaningful outputs. Where archive labels are weak or incomplete, we use scalable synthetic audio generation pipelines that provide strongly labeled examples for detection, event counting, temporal reasoning, and multi-audio in-context learning. Together, these sources cover the same task groups as \benchmark.

We use \dataset{} to train bioacoustic audio-language models and evaluate them on \benchmark{}. Starting from a shared NatureLM-audio base model, we fine-tune a separate model on the \dataset{} supervision for each tier, isolating the effect of each type of supervision to each ability. Tier-specific training substantially improves performance across all tiers of \benchmark{}, including abilities that are weak or absent in existing models. Because of the aligned taxonomy and training recipe, we show that curated metadata improves biological category recognition, LLM-based text synthesis supports richer description and acoustic perception, and synthetic audio enables scene understanding and in-context capabilities.

\definecolor{LightGray}{gray}{0.95}

\tcbset{
  contributionbox/.style={
    enhanced,
    colback=LightGray,
    colframe=PineGreen,
    coltitle=white,
    colbacktitle=PineGreen,
    fonttitle=\bfseries,
    boxrule=0.9pt,
    arc=2pt,
    left=6pt,
    right=6pt,
    top=6pt,
    bottom=6pt,
    titlerule=0pt,
    boxed title style={
      colback=PineGreen,
      colframe=PineGreen,
      arc=2pt,
      boxrule=0pt,
      left=6pt,
      right=6pt,
      top=3pt,
      bottom=3pt
    }
  }
}
\begin{tcolorbox}[contributionbox,title={Outline and Contributions}]
\renewcommand{\thempfootnote}{\arabic{mpfootnote}} %

\begin{enumerate}[leftmargin=*, label=(\arabic*), itemsep=0.45em]

    \item We introduce \textbf{\benchmark}, a benchmark for evaluating bioacoustic audio-language models across \textbf{acoustic perception}, \textbf{biological category recognition}, \textbf{scene understanding}, and \textbf{in-context learning}.

    \item We introduce \textbf{\dataset{}},
    a large-scale bioacoustic audio-language training dataset built from curated archives, underused and audio-derived metadata, LLM-based text synthesis, and synthetic audio pipelines.

    \item We characterize a broader \textbf{bioacoustic ability space} required by ethology and bioacoustics, and show that existing bioacoustic and general-purpose LALMs cover
 only a limited portion of this space.

    \item We demonstrate that training on \textbf{\dataset{}} improves performance across \textbf{\benchmark{}}, including capabilities that are weak or absent in existing models, while also improving performance on established bioacoustic tasks.

    \item We release \textbf{\benchmark} and an evaluation library along with \textbf{\dataset{}} and its synthetic audio subsets%
to support reproducible research on audio-language model evaluation for bioacoustics and ethology.

\end{enumerate}
\end{tcolorbox}

\section{Related Work}\label{sec:related}

With the exception of NatureLM-audio~\citep{robinsonnaturelm}, most prior work on LALMs has focused on human speech, music, and general environmental sound. Models such as LTU, SALMONN, Qwen-Audio, Audio Flamingo, and GAMA connect pretrained audio encoders to large language models in order to support instruction following, question answering, captioning, dialogue, and audio reasoning over broad audio inputs~\citep{gong2023listen, tang2024salmonn, chu2024qwen2audio, ghosh_audio_2026, ghosh2024gama}. Complementary benchmarks such as MMAU evaluate increasingly complex audio understanding and reasoning tasks~\citep{sakshi2024mmau}. However, these models and benchmarks are largely centered on human-centric audio distributions, and general audio competence does not directly imply the fine-grained acoustic perception, biological knowledge, or study-specific adaptability required for bioacoustic research.

Bioacoustics has also seen rapid progress in representation learning. Early work showed that large bird-trained embeddings can transfer beyond birds to other taxa and acoustic domains~\citep{ghani2023global, williams2024leveragingtropicalreefbird}. Self-supervised models such as AVES and animal2vec adapt speech-inspired pretraining objectives to animal vocalizations, improving transfer to downstream bioacoustic classification and detection tasks~\citep{hagiwara2023aves, schaferzimmermann2024animal2vecmeerkatselfsupervisedtransformer}. More recent models, including Bird-MAE and TweetyBERT, further specialize masked prediction objectives for avian soundscapes and fine-grained birdsong parsing~\citep{rauch2025maskedautoencoderslistenbirds, Vengrovski2025}. While these approaches provide strong acoustic representations, they remain primarily encoder-based and are usually evaluated through fixed-output tasks such as species classification, detection, clustering, or individual identification~\citep{schwinger2025foundation, miron2026avex}.
\looseness=-1

Language supervision has only recently been introduced into bioacoustic foundation models. BioLingual uses contrastive audio-language pretraining to support zero-shot classification and retrieval from natural language descriptions~\citep{robinson2024transferable}. NatureLM-audio extends this direction to instruction-following audio-language modeling, enabling zero-shot prediction of species, call type, life stage, and other biologically meaningful labels~\citep{robinsonnaturelm}. These models demonstrate that language-conditioned bioacoustic models are feasible, but their evaluation remains concentrated on biological category recognition. Broader ethological abilities such as acoustic measurement, event counting, call comparison, temporal reasoning, and adaptation to study-specific sound categories remain underexplored.
\looseness=-1

Datasets and benchmarks in bioacoustics have similarly focused on recognition problems. BEANS provides a multi-task benchmark for classification and detection across animal sounds~\citep{hagiwara2023beans}, BIRB evaluates retrieval and generalization in bird vocalizations~\citep{hamer2023birbgeneralizationbenchmarkinformation}, and BirdSet standardizes large-scale avian soundscape classification under domain shift~\citep{rauchbirdset}. Recent audio-language benchmarks have begun to expose the limitations of general LALMs in bioacoustically relevant settings. In particular, WOW-Bench evaluates fine-grained acoustic perception and cognition using marine mammal vocalizations, highlighting the difficulty of low-level listening in unfamiliar acoustic domains~\citep{kim2025wow}. However, existing benchmarks remain limited in taxonomic scope, task structure, or coverage of the broader abilities needed in ethology. In contrast, \benchmark{} targets a broader bioacoustic ability space across a larger variety of taxa. Appendix Table~\ref{tab:benchmark_capability_coverage} compares \benchmark{} with existing benchmarks.

Large archives and datasets such as Xeno-canto, iNaturalist Sounds, and the Animal Sound Archive provide substantial taxonomic breadth, but their annotations are often weak, heterogeneous, or semi-structured~\citep{vellinga2015xeno, chasmai2024inaturalist, AnimalSoundArchiveAccess}. Recent synthetic-data pipelines address part of this limitation by generating strongly labeled examples for sound event detection and few-shot detection~\citep{drasdic, mahon2025robustdetectionoverlappingbioacoustic}. These resources support recognition, detection, and data scaling, but are not organized around the broader capabilities needed for general-purpose bioacoustic audio-language models. \benchmark{} and \dataset{} address this gap with a benchmark and an aligned training corpus built on the same four-tier taxonomy.

\section{Data construction}

\subsection{A taxonomy of bioacoustic audio-language abilities}

While machine learning in bioacoustics has focused primarily on species identification, a broader range of tasks is required by real-world ethology workflows. To structure evaluation and measure progress, we cluster ethological tasks into groups of related tasks we refer to as `tiers'. These tiers align not only with scientific usage but also with categories of capabilities that have been proposed as relevant to develop and evaluate in LALMs~\citep{mammar, starbench, goel_audio_2025}. Each has multiple task families, which are comprised of multiple tasks.

\headinggap
\noindent\textbf{Tier 1: Acoustic perception.}
These tasks measure whether a model can recover the low-level properties of animal sounds, such as fundamental frequency, signal-to-noise ratio (SNR), and acoustic description, without requiring biological category recognition. We instantiate this tier with task families for acoustic-property estimation, acoustic description matching, and acoustic captioning. Scientifically, acoustic measurement is a key part of the workflow of characterizing, interpreting, or comparing acoustic variation~\citep{roadmap}.

\headinggap
\noindent\textbf{Tier 2: Biological category recognition.}
These tasks ask what biological or ecological categories are present, such as species, broader taxonomic groups, call types, life stage, or behavior. We instantiate this tier with task families for species and taxonomy recognition, taxon presence detection, call-type recognition, life-stage and behavior prediction, and semantic captioning. These are closest to established bioacoustic benchmarks~\citep{hagiwara2023beans, rauchbirdset, robinsonnaturelm} and support biodiversity monitoring, repertoire analysis, and other analyses of communication.

\headinggap
\noindent\textbf{Tier 3: Scene understanding.}
These tasks measure whether a model can interpret longer, structured recordings containing multiple vocalizations, species, or individuals. We instantiate this tier with task families for event and species counting, temporal localization, co-occurrence, order reasoning, acoustic comparison across events, species listing, and structured scene description. This setting is especially relevant for passive acoustic monitoring, where recordings often contain overlapping calls, repeated events, multiple individuals or species, and long stretches of background sound. Scientifically, these tasks support questions about not only what occurred, but also when.

\headinggap
\noindent\textbf{Tier 4: In-context learning.}
These tasks measure whether a model can adapt to a task specified at inference time using text descriptions or audio examples. We instantiate this tier with task families for audio-example-conditioned classification, few-shot detection, unseen-species recognition, and call-type matching. For example, a researcher may provide examples of an unknown call type, a rare species, or a project-specific sound category and ask the model to identify matching sounds in new recordings. Because bioacoustic data are long-tailed across taxa, tasks, and recording conditions, in-context abilities are critical for learning new tasks at runtime without training a separate classifier for every study.

\subsection{Text creation methods}

Both \benchmark{} and \dataset{} require converting bioacoustic annotations into instruction-following audio-language examples. We use two complementary text creation methods: templated construction and LLM-based construction. These are shared across the benchmark and the dataset. However, \benchmark{} is exclusively grounded on human-annotated metadata, while \dataset{} scales supervision by additionally relying on audio-derived metadata, such as acoustic measurements and spectrogram-based descriptions, and synthetic audio.

\headinggap
\noindent\textbf{Templated construction}

We construct prompt-response pairs from structured fields such as species identity, taxonomy, locality, call type, life stage, and other available metadata. Each source dataset is mapped into a unified schema, transformed in a task-dependent way, and then converted into multiple task variants using prompt and response templates and algorithmic transformations. The full set of templated tasks and source datasets is provided in Appendix~\ref{sec:template-construction}.

\headinggap
\noindent\textbf{LLM-based construction}\label{sec:llm-based-construction}

Templated construction provides reliable supervision when metadata are already structured, but it cannot capture the full range of information available in bioacoustic archives. Many recordings are also accompanied by free-text or semi-structured annotations, such as field notes, that contain scientifically useful information but are difficult to convert into fixed templates without losing nuance. We therefore use an LLM-based pipeline to transform such evidence into natural-language question-answer pairs and descriptions (overview in Appendix~\ref{llm_overview}).

For each task, we write a dedicated prompt specifying the available evidence, the intended task, the output format, constraints on what information may be used, and skip conditions for when the evidence is insufficient or ambiguous. Compared with broad multi-task generation prompts~\citep{ghosh_audio_2026}, this single-task design gives finer control over task definitions, their alignment with bioacoustic use cases, and their relative frequency, and narrows each generation problem enough to use a smaller, faster LLM, Gemma 4 26B MoE~\citep{gemma4}. Where acoustic supervision is scarce, we also use Gemma's vision capabilities to describe vocalizations characteristics from spectrograms, prompting with human-written examples and restricting descriptions to acoustic properties.

Generated text may contain unsupported claims, inaudible information, references to information that should not be available to the model, or other errors. We therefore filter outputs with regex-based checks for format, keyword, and schema compliance, and with LLM-as-a-judge scores for instruction following with respect to ground truth, hallucination, audibility, and difficulty. One author manually reviewed samples of pre- and post-filtered outputs for each task category as a quality-control step. LLM-based tasks and source datasets are listed in Appendix~\ref{sec:synthetic-construction}.

\subsection{\benchmark{} benchmark construction}

BEANS-Next is constructed as a held-out evaluation suite aligned with the four-tier taxonomy in Section 3.1. Evaluation examples are built from real recordings and human- or dataset-provided annotations. Templated and LLM-based text construction are used only to convert supported annotations into prompts and responses, not to introduce new labels.
All BEANS-Next benchmark recordings, and any derivatives of those recordings, are strictly disjoint from ROOTS.
For tasks where species identity is central, we additionally follow the held-out species protocol introduced in NatureLM-audio~\citep{robinsonnaturelm}: BEANS-Next uses the same held-out species split where applicable, excluding those taxa from the corresponding ROOTS/NatureLM-audio training mixture. We list the held-out taxa in Appendix~\ref{sec:held-out}.
Per-tier data sources, task formats, sample counts, and metrics are detailed in Appendix~\ref{sec:benchmark-creation}.

\subsection{\dataset{} training data construction}
\dataset{} is an instruction-following training corpus built from curated bioacoustic archives, task-specific annotated datasets, audio-derived metadata, and the synthetic audio pipelines described below. Each source is mapped into a shared schema. Structured fields such as species, taxonomy, call type, life stage, sex, and locality are converted with templates, while free-text notes, acoustic measurements, and spectrogram-derived descriptions are converted with the LLM-based pipeline (Section~\ref{sec:llm-based-construction}). Together, these sources provide supervision across all four tiers of the taxonomy. Source-specific tasks and counts are given in Appendices~\ref{roots_details} and~\ref{sec:corpus}.

\headinggap
\noindent\textbf{Synthetic Audio Pipelines}

For most species, bioacoustic data with temporally strong labels are unavailable~\citep{nolasco2023learning}. In addition, archives used to train bioacoustic foundation models often contain label noise and missing labels~\citep{michaud2023unsupervised}. To address these gaps, we use heavy augmentation pipelines to synthesize strongly labeled bioacoustic recordings at scale, which are then passed to the text construction pipelines. We summarize the resulting datasets here, with details in Appendix~\ref{sec:synthetic-audio}.

We construct two single-audio synthetic datasets. \textbf{Strong detection} converts weakly labeled bioacoustic audio into synthetic, strongly labeled recordings with precise timestamps for each species. These recordings are used by the text synthesis pipeline to generate scene understanding tasks. \textbf{Detect-diarize} creates synthetic recordings with temporally strong labels and a proxy speaker ID for each call. Starting from BirdMixIT source-separated tracks~\citep{denton2022improving}, we extract call events and add them to synthetic scenes while preserving their original order and timing and placing them at a consistent amplitude scale.

We also construct synthetic multi-audio datasets. \textbf{Few-shot multi-label classification} gives audio examples of up to four target classes and the model must predict which of those classes are present in a query audio clip. \textbf{Call-type tasks} use sound clusters to construct synthetic call-type tasks in multiple formats, including multiple choice, binary presence prediction, and same--different prediction between two sounds. Positive examples are sampled from the same species and acoustic cluster according to our processing pipeline, while negatives are sampled from random clusters outside the target cluster.

\section{Experimental Setup}

\begin{table}[t]
\centering
\small
\renewcommand{\arraystretch}{1.0}
\caption{BEANS-Zero classification results (accuracy). Higher is better.}
\label{tab:beans-zero}
\begin{tabular}{l c c c c c c}
\toprule
Model & CBI & HBDB & ESC50 & Call-type & Lifestage & N-indv \\
\midrule
AF-Next              & 0.006 & 0.004 & 0.768 & 0.078 & 0.048 & 0.579 \\
Qwen3-Omni           & 0.040 & 0.038 & 0.797 & 0.018 & 0.054 & 0.531 \\
NatureLM-audio v1.0  & 0.778 & \textbf{0.114} & \textbf{0.820} & 0.870 & \textbf{0.794} & \textbf{0.658} \\
\textit{\model}      & \textbf{0.808} & 0.062 & 0.360 & \textbf{0.882} & 0.753 & 0.501 \\
\bottomrule
\end{tabular}%

\end{table}

\begin{table}[t]
\centering
\small
\renewcommand{\arraystretch}{1.0}
\caption{BirdSet results (top-1 accuracy). Higher is better.}
\label{tab:birdset}
\begin{tabular}{l c c c c c c c c}
\toprule
Model & HSN & NBP & NES & PER & POW & SNE & UHH & AVG \\
\midrule

AF-Next             & 0.000 & 0.000 & 0.000 & 0.000 & 0.002 & 0.002 & 0.000 & 0.001 \\
Qwen3-Omni          & 0.003 & 0.014 & 0.000 & 0.000 & 0.067 & 0.000 & 0.000 & 0.012 \\
NatureLM-audio v1.0 & 0.273 & 0.601 & 0.435 & \textbf{0.557} & 0.891 & 0.656 & 0.339 & 0.536 \\
\textit{\model}     & \textbf{0.325} & \textbf{0.644} & \textbf{0.446} & 0.503 & \textbf{0.953} & \textbf{0.715} & \textbf{0.404} & \textbf{0.570} \\
\bottomrule
\end{tabular}
\end{table}

\subsection{Models}

We evaluate the bioacoustic audio-language model NatureLM-audio v1.0~\citep{robinsonnaturelm} and two strong open general-purpose LALMs, Audio Flamingo Next (AF-Next)~\citep{ghosh_audio_2026} and Qwen3-Omni~\citep{xu2025qwen3}. We focus on generative LALMs because most \benchmark{} tasks require open-ended or multi-audio answers that encoder-only and contrastive models cannot produce; proprietary API models were excluded due to high failure rates and cost (Appendix~\ref{sec:evaluation}). We additionally train NatureLM-audio v1.1 as the base model for our \dataset{} experiments; it retains the v1.0 architecture but adds expanded training data and support for longer and multiple audio inputs. All models are queried through the same inference, parsing, and scoring pipeline.

\subsection{Training}

NatureLM-audio v1.1 is trained on an expanded version of the NatureLM-audio v1.0 semantic task mixture, using curated metadata and primarily templated supervision. From this checkpoint, we fine-tune a separate model on the additional \dataset{} supervision for each tier (NatureLM-audio T1--T4), with the audio encoder and Q-former unfrozen and the LLM fine-tuned with LoRA~\citep{lora} (Appendix~\ref{sec:ft_hypers}). Holding architecture and initialization fixed isolates the effect of the training data: because the base model has seen primarily templated semantic supervision, comparisons against it measure the contribution of LLM-based text synthesis, synthetic audio, and audio-derived metadata. We do not train a single joint model in this work; joint multi-tier training, possible interference between tiers, and architectural changes are left to future work.

\section{Results}
\providecommand{\prate}[1]{\,{\color{black!55}\scriptsize #1\%}}
\providecommand{\lowpr}[1]{{\color{black!55}#1}}
\begin{table}[t]
\centering
\caption{BEANS-Next Tier 1 (acoustic perception) results. Captioning uses CIDEr ($\times 100$); description matching and SNR MCQ use accuracy (\%); SNR regression and mean $f_0$ use MAE in dB and Hz, respectively. Higher is better unless marked $\downarrow$. Small grey percentages give the parse rate (share of answers with a parseable number); MAE is computed over parsed answers only, and grey MAEs rest on fewer than half of the answers and are not bolded.}
\small
\label{tab:acoustic}
\renewcommand{\arraystretch}{1.0}
\setlength{\tabcolsep}{3.2pt}
\begin{tabular}{@{}lrrrrrrr@{}}
\toprule
& \multicolumn{3}{c}{Acoustic description}
& \multicolumn{2}{c}{SNR}
& \multicolumn{2}{c}{Mean $f_0$ MAE (Hz) $\downarrow$} \\
\cmidrule(lr){2-4}
\cmidrule(lr){5-6}
\cmidrule(lr){7-8}
Model & Caption & Desc. MCQ & Crow & MCQ & MAE (dB) $\downarrow$ & Seen & Held-out \\
\midrule
AF-Next & 0.5 & 50.5 & \textbf{48.0} & 36.7 & \lowpr{31.0}\prate{1.5} & \lowpr{5706}\prate{13.7} & \lowpr{268}\prate{1.6} \\
Qwen3-Omni & 0.0 & 43.6 & 47.0 & 25.2 & 15.4\prate{94.5} & 3749\prate{99.9} & 208\prate{100} \\
NatureLM-audio v1.0 & 0.3 & 28.4 & 33.5 & 34.6 & 22.6\prate{92.2} & 4086\prate{100} & 208\prate{100} \\
\textit{\model} & 0.0 & 28.7 & 30.0 & 26.7 & 19.1\prate{52.1} & 4334\prate{84.0} & 360\prate{91.1} \\
\textit{NatureLM-audio T1} & \textbf{2.8} & \textbf{97.9} & 27.5 & \textbf{60.8} & \textbf{10.4}\prate{100} & \textbf{253}\prate{100} & \textbf{125}\prate{100} \\
\bottomrule
\end{tabular}
\end{table}

\begin{table}[t]
\centering
\caption{BEANS-Next Tier 2 (biological category recognition) results. Presence and behavior use accuracy (\%), fixed-vocabulary call type uses macro-F1 (\%) averaged over the five labels, and captioning uses CIDEr ($\times 100$). Higher is better.}
\small
\renewcommand{\arraystretch}{1.0}
\label{tab:semantic}
\setlength{\tabcolsep}{3pt}
\begin{tabular}{@{}lrrrrrrrrrr@{}}
\toprule
& \multicolumn{4}{c}{Taxon presence}
& \multicolumn{3}{c}{Call-type presence}
& & & \\
\cmidrule(lr){2-5}
\cmidrule(lr){6-8}
Model & Bird & Mamm. & Amph. & Insect & Alarm & Flight & Beg. & \shortstack{Call\\type} & Behav. & Caption \\
\midrule
AF-Next & 71.2 & 67.1 & 81.2 & 75.5 & 56.2 & 46.5 & 54.9 & 44.4 & 56.5 & 0.1 \\
Qwen3-Omni & 68.4 & 65.0 & 85.7 & 73.4 & 66.8 & 57.0 & 65.7 & 25.6 & 56.7 & 0.1 \\
NatureLM-audio v1.0 & 55.7 & 63.4 & 68.0 & 56.0 & 50.3 & 51.7 & 50.2 & 48.2 & 39.9 & \textbf{1.7} \\
\textit{\model} & 54.7 & 65.2 & 53.2 & 59.3 & 58.8 & 71.5 & 62.5 & 39.5 & 33.7 & 1.1 \\
\textit{NatureLM-audio T2} & \textbf{91.5} & \textbf{87.9} & \textbf{96.4} & \textbf{93.4} & \textbf{69.9} & \textbf{82.2} & \textbf{74.6} & \textbf{59.0} & \textbf{87.1} & 1.2 \\
\bottomrule
\end{tabular}
\label{tab:beanspro-t2-results}
\end{table}

On the established benchmarks (Tables~\ref{tab:beans-zero} and~\ref{tab:birdset}), NatureLM-audio models are far stronger than general-purpose LALMs at species-level classification. On BirdSet, Audio Flamingo Next and Qwen3-Omni score near zero, mostly answering `Bird' or `None'. NatureLM-audio v1.1 improves over v1.0 on CBI, call type, and BirdSet, but regresses on ESC-50, HumBugDB, life stage, and individual counting.

\subsection{Tier 1: Acoustic Perception}
\providecommand{\prate}[1]{\,{\color{black!55}\scriptsize #1\%}}
\providecommand{\lowpr}[1]{{\color{black!55}#1}}
\begin{table}[t]
\centering
\caption{BEANS-Next Tier 3 (scene understanding) results. Classification and temporal tasks use top-1 accuracy (\%); counting tasks use mean absolute error (MAE); the frequency-range task (Range) uses coverage-aware species-band IoU (\%, mean over ground-truth species, missed species scored 0); summary and captioning use corpus CIDEr ($\times 100$). \texttt{S}: species, \texttt{V}: vocalization, \texttt{\#}: number of. Higher is better unless marked $\downarrow$. Small grey percentages give the parse rate (share of answers with a parseable number); MAE is computed over parsed answers only, and grey MAEs rest on fewer than half of the answers and are not bolded.}
\label{tab:beansnext-t3-results}

\small
\renewcommand{\arraystretch}{1.0}

\setlength{\tabcolsep}{2.4pt}
\begin{tabular}{@{}lrrrrrrrr@{}}
\toprule
& \multicolumn{3}{c}{Counting MAE $\downarrow$}
& \multicolumn{2}{c}{Temporal}
& \multicolumn{3}{c}{Scene / open} \\
\cmidrule(lr){2-4}
\cmidrule(lr){5-6}
\cmidrule(lr){7-9}
Model & \texttt{\#S} & \texttt{\#V} & \texttt{\#V}/\texttt{S} & Referring & Overlap & Range & Sum. & Caption \\
\midrule
AF-Next & 1.64\prate{97.2} & 4.26\prate{99.6} & 2.15\prate{52.5} & 33.5 & 47.0 & 0.0 & 0.0 & 0.0 \\
Qwen3-Omni & \textbf{1.38}\prate{97.0} & 3.93\prate{99.9} & 1.94\prate{88.1} & 25.0 & 46.3 & 0.0 & 0.0 & 0.0 \\
NatureLM-audio v1.0 & 1.54\prate{86.1} & 4.09\prate{100} & \lowpr{1.94}\prate{1.6} & 28.0 & 59.0 & 0.0 & 0.1 & 0.0 \\
\textit{\model} & 1.50\prate{76.7} & 4.57\prate{93.4} & \lowpr{1.49}\prate{12.8} & 30.0 & 47.1 & 0.0 & 2.4 & 0.1 \\
\textit{NatureLM-audio T3} & 1.78\prate{100} & \textbf{3.81}\prate{100} & \textbf{1.72}\prate{99.9} & \textbf{48.0} & \textbf{61.7} & \textbf{11.3} & \textbf{5.0} & \textbf{1.6} \\
\bottomrule
\end{tabular}

\vspace{0.6em}

\setlength{\tabcolsep}{4pt}
\begin{tabular}{lrrrrrrrrrr}
\toprule
& \multicolumn{10}{c}{Species ID by characteristic} \\
\cmidrule(lr){2-11}
& \multicolumn{5}{c}{Open-ended}
& \multicolumn{5}{c}{MCQ} \\
\cmidrule(lr){2-6}
\cmidrule(lr){7-11}
Model
& Order
& High
& Low
& Long
& Freq.
& Order
& High
& Low
& Long
& Freq. \\
\midrule
AF-Next
& 0.4
& 3.8
& 0.0
& 5.3
& 0.0
& 31.4
& 35.0
& 50.3
& 35.6
& 30.5 \\
Qwen3-Omni
& 1.0
& 0.5
& 0.2
& 0.4
& 0.6
& 36.1
& 49.5
& 48.8
& 43.5
& 34.8 \\
NatureLM-audio v1.0
& 34.7
& 15.9
& 13.8
& 14.2
& 27.0
& 37.7
& 43.4
& 36.2
& 43.5
& 40.7 \\
\textit{\model}
& 30.6
& 33.3
& 25.9
& \textbf{37.3}
& 36.0
& 29.2
& 31.6
& 32.0
& 29.5
& 30.1 \\
\textit{NatureLM-audio T3}
& \textbf{41.9}
& \textbf{37.0}
& \textbf{32.0}
& 33.2
& \textbf{38.7}
& \textbf{47.2}
& \textbf{61.8}
& \textbf{58.1}
& \textbf{48.6}
& \textbf{51.1} \\
\bottomrule
\end{tabular}

\end{table}

Table~\ref{tab:acoustic} shows that \dataset{} Tier 1 training adds signal-level capabilities that are weak in the NatureLM-audio v1.1 base model, improving acoustic descriptions, SNR estimation, and mean $f_0$ prediction. The improvement on unseen-species $f_0$ estimation is particularly important because it suggests that the model is not only exploiting taxon-specific priors, but has learned a more transferable mapping from audio to quantitative acoustic attributes. The main exception is crow vocalization description, where the general-purpose LALMs remain better;
 this likely points to residual sensitivity to domain and annotation format, rather than uniform transfer across all acoustic-description settings.

\subsection{Tier 2: Biological Category Recognition}
\begin{table}[t]
\centering
\caption{BEANS-Next Tier 4 (in-context learning) results. Gibbons and DCASE use macro-F1 averaged over reference labels (A--C and A--D, respectively), with no detection represented by an empty set; other tasks use accuracy. Scores are on a 0--1 scale; higher is better. NatureLM-audio v1.0 does not accept multiple audio inputs.}
\small
\renewcommand{\arraystretch}{1.0}
\label{tab:multi-audio-fewshot}
\setlength{\tabcolsep}{5pt}
\begin{tabular}{lccccc}
\toprule
& \multicolumn{5}{c}{Few-shot task} \\
\cmidrule(lr){2-6}
Model
& Gibbons
& Otters
& DCASE
& Crows
& Unseen \\
\midrule
AF-Next
& 0.007
& 0.255
& 0.119
& 0.225
& 0.230 \\
Qwen3-Omni
& 0.175
& 0.535
& 0.327
& 0.620
& 0.480 \\
\textit{NatureLM-audio v1.1}
& 0.176
& 0.200
& 0.048
& 0.240
& 0.270 \\
\textit{NatureLM-audio T4}
& \textbf{0.243}
& \textbf{0.645}
& \textbf{0.452}
& \textbf{0.715}
& \textbf{0.650} \\
\bottomrule
\end{tabular}
\end{table}

Table~\ref{tab:semantic} shows that \dataset{} Tier 2 training also improves biological category recognition. The gains across presence, call-type, and behavior prediction suggest that mined metadata provides useful supervision beyond species labels alone. This supports the broader claim that improving bioacoustic LALMs is not only a matter of adding more labeled recordings, but also of expanding the kinds of biological supervision available to the model. Captioning remains weak in absolute terms, indicating that open-ended biological description remains harder than discriminative metadata prediction.

\subsection{Tier 3: Scene Understanding}

Table~\ref{tab:beansnext-t3-results} shows that \dataset{} Tier 3 training mainly improves the parsing of scenes into events and relations. The largest gains occur when species identities must be bound to event-level attributes, such as which species vocalized first, which vocalization was highest or lowest in pitch, and whether vocalizations overlapped, with consistent gains across all multiple-choice characteristic tasks. Because the synthetic scenes explicitly control species identity, timing, overlap, duration, and frequency range, they provide direct supervision for this kind of structured reasoning. The supervision is nonetheless incomplete: counting remains difficult, with a species-count error slightly worse than the base models, and open-ended duration-based identification does not improve. One hypothesis is that the architecture limits fine-grained temporal perception, consistent with recent moves toward denser audio embeddings and timestamp-based positional encodings~\citep{ghosh_audio_2026}.

\subsection{Tier 4: In-Context Learning}

Table~\ref{tab:multi-audio-fewshot} shows the strongest evidence that \dataset{} can change how the model uses its inputs. Audio Flamingo Next and the NatureLM-audio v1.1 base model remain near chance on the matching tasks, suggesting that they do not use the support audio as task-defining context. Qwen3-Omni performs well above chance on call-type and unseen-species matching, indicating that general-domain in-context learning partially transfers to bioacoustics. Tier 4 training yields the best results on all five tasks, showing that synthetic multi-audio supervision teaches the model to compare a query against in-context examples rather than rely only on labels seen in training. Gains are smallest on Gibbons, a low-SNR dataset, suggesting limited robustness to recording conditions and to the domain shift from synthetic data.

\section{Conclusion}

Existing audio-language models, both general-purpose and bioacoustic, cover only a limited portion of the capabilities ethologists and bioacousticians actually rely on. By organizing evaluation and training around a shared capability taxonomy, we are able to measure these gaps and show that targeted data construction---curated metadata, LLM-based text synthesis, and synthetic audio pipelines---can begin to close them. We see this as a step toward audio-language models that support the full range of bioacoustic workflows rather than only naming what is present. Substantial headroom remains, and we hope the released benchmark and dataset help the community make progress on these.

These capabilities also raise concerns worth flagging. Both BEANS-Next and ROOTS inherit taxonomic and geographic biases from public bioacoustic archives, which may limit reliability for underrepresented taxa and ecosystems. Stronger detection of endangered or vulnerable species could be misused when combined with geolocation metadata, a risk we partially mitigate by excluding endangered species from our Xeno-canto sources. And as audio-language models are deployed more widely in ecological workflows, errors—including hallucinated species, behaviors, or counts—could propagate into downstream scientific or conservation decisions. We therefore recommend these models be used as assistive tools, with predictions validated by domain experts.

\subsection*{AI use statement}

We used generative AI tools (Gemma 4 26B MoE) as a core component of our data construction pipelines, specifically synthesizing question-answer pairs and descriptions from semi-structured and unstructured metadata, generating spectrogram-derived acoustic descriptions, and scoring candidate examples via an LLM-as-a-judge filtering step (instruction following, hallucination, audibility, and difficulty), as described in Section~\ref{sec:llm-based-construction} and Appendix~\ref{sec:synthetic-construction}. One of the authors manually reviewed a sample of pre- and post-filtered LLM outputs for each category of data as an additional quality-control step.

We also used AI coding assistants to help implement portions of the training, evaluation, data processing, and data synthesis code, and to assist with routine cluster job management and orchestration. All code was reviewed and tested by at least one of the authors before use in reported experiments.

Generative AI was used to refine some of the writing in the paper, after which it was carefully reviewed for accuracy. It was also used to spot any obvious omissions or mistakes before submission, to generate visual elements for Figure \ref{fig:overview} and \ref{fig:pipeline}, and to edit the format of tables, in which case a manual review of table values was conducted. We take responsibility for the final content of this work, including text, claims, and artifacts produced with the aid of generative AI.

\subsection*{Reproducibility statement}

We release the following artifacts to support the reproducibility of our results. \href{https://huggingface.co/datasets/EarthSpeciesProject/BEANS-Next}{\benchmark{}} and \href{https://huggingface.co/datasets/EarthSpeciesProject/ROOTS}{\dataset{}} are publicly available on Hugging Face, together with the synthetic audio subsets \href{https://huggingface.co/datasets/EarthSpeciesProject/synthetic-strong-detection}{Synthetic Strong Detection}, \href{https://huggingface.co/datasets/EarthSpeciesProject/synthetic-detect-diarize}{Synthetic Detect-Diarize}, and \href{https://huggingface.co/datasets/EarthSpeciesProject/animalspeak-pseudovox}{AnimalSpeak-PseudoVox}. For \dataset{}, we do not redistribute audio. However, all audio used is publicly available, and we provide guidance for retrieving and preprocessing it accordingly. The \href{https://github.com/earthspecies/beans-next}{\benchmark{} evaluation library} is publicly available on GitHub and implements data loading, model-server interfaces, answer parsing, and metric computation. Benchmark construction, task definitions, sample counts, metrics, and held-out taxa are described in Appendix~\ref{sec:benchmark-creation}, including Table~\ref{sec:evals} and Appendix~\ref{sec:held-out}. \dataset{} construction, covering templated and LLM-based tasks, filtering, and synthetic audio generation, is described in Appendix~\ref{roots_details}. Templates are provided on Hugging Face with \dataset{}, as well as their mapping to specific samples. Training and fine-tuning hyperparameters, augmentations, and compute are given in Appendices~\ref{sec:hypers} and~\ref{sec:ft_hypers}. The evaluation protocol, post-processing, and metric definitions are in Appendix~\ref{sec:evaluation}, and the text-only and noise controls are in Appendix~\ref{sec:text-only-baselines}.

\clearpage
\printbibliography

\medskip

\clearpage
\appendix

\section{Appendix}

In this Appendix, we provide additional details on training and evaluation.

\section{\dataset{} Construction}\label{roots_details}
\subsection{Templated Data Construction} \label{sec:template-construction}

\begin{longtable}{@{}p{2.8cm} p{4.8cm} p{4.9cm}@{}}
\caption{Datasets used for each training task, with a representative prompt.}
\label{tab:datasets-by-task} \\
\toprule
\textbf{Task} & \textbf{Example Prompt} & \textbf{Datasets} \\
\midrule
\endfirsthead
\toprule
\textbf{Task} & \textbf{Example Prompt} & \textbf{Datasets} \\
\midrule
\endhead
\midrule
\multicolumn{3}{r@{}}{\footnotesize\itshape Continued on next page} \\
\endfoot
\bottomrule
\endlastfoot
  \footnotesize Species ID (common name) & \footnotesize What species is vocalizing in this audio recording? Common name? & \footnotesize Xeno-Canto, iNaturalist, Watkins, Animal Sound Archive \\
  \footnotesize Species ID (scientific name) & \footnotesize What is the scientific name of the focal species in the audio? & \footnotesize Xeno-Canto, iNaturalist, Watkins, Animal Sound Archive \\
  \footnotesize Species ID + context (common) & \footnotesize Given the context: `\{\{context\}\}', what is the common name for the focal species? & \footnotesize Xeno-Canto, iNaturalist, Animal Sound Archive \\
  \footnotesize Species ID + context (scientific) & \footnotesize Given the context: `\{\{context\}\}', what is the scientific name for the focal species? & \footnotesize Xeno-Canto, iNaturalist, Animal Sound Archive \\
  \footnotesize Full taxonomy + context & \footnotesize Given the context: `\{\{context\}\}', provide the full taxonomic classification. & \footnotesize Xeno-Canto, iNaturalist, Animal Sound Archive \\
  \footnotesize Genus classification & \footnotesize What is the genus of the focal species in the audio? & \footnotesize Xeno-Canto, iNaturalist, Animal Sound Archive \\
  \footnotesize Family classification & \footnotesize What is the family of the focal species in the audio? & \footnotesize Xeno-Canto, iNaturalist, Animal Sound Archive \\
  \footnotesize Order classification & \footnotesize What is the order of the focal species in the audio? & \footnotesize Xeno-Canto, iNaturalist, Animal Sound Archive \\
  \footnotesize Full taxonomy classification & \footnotesize Provide the full taxonomic classification for the species, starting with the phylum. & \footnotesize Xeno-Canto, iNaturalist, Watkins, Animal Sound Archive \\
  \footnotesize Call type classification & \footnotesize What type of vocalization or call is this? & \footnotesize Xeno-Canto, Watkins, Animal Sound Archive \\
  \footnotesize Call type (species-conditioned) & \footnotesize What type of call is the \{\{species\_common\}\} making? & \footnotesize Xeno-Canto, Watkins, Animal Sound Archive \\
  \footnotesize Call vs.~song (binary) & \footnotesize Is this a call or a song? & \footnotesize Xeno-Canto \\
  \footnotesize Alarm call presence (binary) & \footnotesize Is an alarm call present in this recording? Answer Yes or No. & \footnotesize Xeno-Canto, Animal Sound Archive \\
  \footnotesize Alarm call + species (binary) & \footnotesize Is the \{\{species\_common\}\} making an alarm call? Yes or No. & \footnotesize Xeno-Canto, Animal Sound Archive \\
  \footnotesize Alarm / other call / song (3-way) & \footnotesize Classify this recording as one of: alarm, other\_call, song. & \footnotesize Xeno-Canto, Animal Sound Archive \\
  \footnotesize Call types present (multi-label) & \footnotesize List all call types present as a comma-separated list. & \footnotesize Xeno-Canto, Animal Sound Archive \\
  \footnotesize Call type (fixed 5-label vocab) & \footnotesize Which are present? alarm call, flight call, begging call, song, call. & \footnotesize Xeno-Canto, Animal Sound Archive \\
  \footnotesize Fine-grained call type & \footnotesize What are the fine-grained vocalization types in this recording? & \footnotesize Watkins \\
  \footnotesize Fine call type (species-cond.) & \footnotesize What fine-grained vocalization types is the \{\{species\_common\}\} producing? & \footnotesize Watkins \\
  \footnotesize Life stage classification & \footnotesize What life stage is the animal in this recording? & \footnotesize Xeno-Canto, iNaturalist, Animal Sound Archive \\
  \footnotesize Life stage (species-conditioned) & \footnotesize What is the life stage of the \{\{species\_common\}\}? & \footnotesize Xeno-Canto, iNaturalist, Animal Sound Archive \\
  \footnotesize Adult vs.~juvenile (binary) & \footnotesize Is the focal species an adult or juvenile? & \footnotesize Xeno-Canto, iNaturalist \\
  \footnotesize Sex prediction & \footnotesize Is the animal vocalizing male or female? & \footnotesize Xeno-Canto, Animal Sound Archive \\
  \footnotesize Sex prediction (species-cond.) & \footnotesize What is the sex of the \{\{species\_common\}\}? & \footnotesize Xeno-Canto, Animal Sound Archive \\
  \footnotesize Species + behavior (common) & \footnotesize What species is vocalizing and what type of vocalization? & \footnotesize Xeno-Canto, Watkins, Animal Sound Archive \\
  \footnotesize Species + behavior (scientific) & \footnotesize Give the scientific name and vocalization type. & \footnotesize Xeno-Canto, Animal Sound Archive \\
  \footnotesize Species + fine call type & \footnotesize Identify the species and list the detailed vocalization types. & \footnotesize Watkins \\
  \footnotesize Species + life stage (common) & \footnotesize What species is in this recording and what is its life stage? & \footnotesize Xeno-Canto, iNaturalist, Animal Sound Archive \\
  \footnotesize Species + life stage (scientific) & \footnotesize Give the scientific name and life stage. & \footnotesize Xeno-Canto, iNaturalist, Animal Sound Archive \\
  \footnotesize Species + behavior + life stage & \footnotesize Identify the species, call type, and life stage. & \footnotesize Xeno-Canto, Animal Sound Archive \\
  \footnotesize Open detection (common name) & \footnotesize Identify any species present by their common name. & \footnotesize Xeno-Canto, iNaturalist, Watkins, Animal Sound Archive \\
  \footnotesize Open detection (scientific name) & \footnotesize Identify any species present by their scientific name. & \footnotesize Xeno-Canto, iNaturalist, Watkins, Animal Sound Archive \\
  \footnotesize Open detection (full taxonomy) & \footnotesize Provide the full taxonomic classification for any species heard. & \footnotesize Xeno-Canto, iNaturalist, Watkins, Animal Sound Archive \\
  \footnotesize Species options (common, classif.) & \footnotesize Which of these is the focal species? Options: \{\{species\_choices\}\} & \footnotesize Xeno-Canto, iNaturalist, Watkins, Animal Sound Archive \\
  \footnotesize Species options (common, detect.) & \footnotesize Which of these species, if any, are present? \{\{species\_choices\}\} & \footnotesize Xeno-Canto, iNaturalist, Watkins, Animal Sound Archive \\
  \footnotesize Species options (scientific, classif.) & \footnotesize Which of these species (scientific name) is in the audio? \{\{species\_choices\}\} & \footnotesize Xeno-Canto, iNaturalist, Watkins, Animal Sound Archive \\
  \footnotesize Species options (scientific, detect.) & \footnotesize Which of these species, if any, are present? \{\{species\_choices\}\} & \footnotesize Xeno-Canto, iNaturalist, Watkins, Animal Sound Archive \\
  \footnotesize Context + options (common, classif.) & \footnotesize Context: \{\{context\}\}. Which species? \{\{species\_choices\}\} & \footnotesize Xeno-Canto, iNaturalist \\
  \footnotesize Context + options (common, detect.) & \footnotesize Context: \{\{context\}\}. Which, if any? \{\{species\_choices\}\} & \footnotesize Xeno-Canto, iNaturalist \\
  \footnotesize Context + options (scientific, classif.) & \footnotesize Context: \{\{context\}\}. Which species (scientific)? \{\{species\_choices\}\} & \footnotesize Xeno-Canto, iNaturalist \\
  \footnotesize Context + options (scientific, detect.) & \footnotesize Context: \{\{context\}\}. Which, if any (scientific)? \{\{species\_choices\}\} & \footnotesize Xeno-Canto, iNaturalist \\
  \footnotesize Genus options (classification) & \footnotesize Which of these genera matches the species? \{\{species\_choices\}\} & \footnotesize Xeno-Canto, iNaturalist, Animal Sound Archive \\
  \footnotesize Genus options (detection) & \footnotesize Which of these genera, if any, matches? \{\{species\_choices\}\} & \footnotesize Xeno-Canto, iNaturalist, Animal Sound Archive \\
  \footnotesize Family options (classification) & \footnotesize Which of these families matches the species? \{\{species\_choices\}\} & \footnotesize Xeno-Canto, iNaturalist, Animal Sound Archive \\
  \footnotesize Family options (detection) & \footnotesize Which of these families, if any, matches? \{\{species\_choices\}\} & \footnotesize Xeno-Canto, iNaturalist, Animal Sound Archive \\
  \footnotesize Taxonomic options (classification) & \footnotesize Which taxonomic classification matches? \{\{species\_choices\}\} & \footnotesize Xeno-Canto, iNaturalist, Animal Sound Archive \\
  \footnotesize Taxonomic options (detection) & \footnotesize Which classification, if any, matches? \{\{species\_choices\}\} & \footnotesize Xeno-Canto, iNaturalist, Animal Sound Archive \\
  \footnotesize Multi-label species (scientific) & \footnotesize List the scientific names of all species vocalizing in this clip. & \footnotesize Xeno-Canto \\
  \footnotesize Multi-label species (common) & \footnotesize List the common names of all species vocalizing in this clip. & \footnotesize Xeno-Canto \\
  \footnotesize Multi-label species + count & \footnotesize How many species are vocalizing, and what are they (scientific)? & \footnotesize WABAD, SSW \\
  \footnotesize Multi-label species (soundscape) & \footnotesize List the scientific names of all species in this soundscape. & \footnotesize WABAD, SSW \\
  \footnotesize Audio captioning (general) & \footnotesize Caption this audio with a rich, detailed description. & \footnotesize AudioSet \\
  \footnotesize Bioacoustic captioning (XC synth.) & \footnotesize Caption the audio, using common names for any animal species. & \footnotesize Xeno-Canto \\
  \footnotesize Bioacoustic captioning (iNat synth.) & \footnotesize Caption the audio, using common names for any animal species. & \footnotesize iNaturalist \\
  \footnotesize Bioacoustic captioning (iNat short) & \footnotesize Give a short caption for this recording, using common names. & \footnotesize iNaturalist \\
  \footnotesize Bioacoustic captioning (iNat sci.) & \footnotesize Give a short caption, using scientific names for any animal species. & \footnotesize iNaturalist \\
  \footnotesize Call count per species & \footnotesize How many calls from each species? Give scientific names. & \footnotesize WABAD \\
  \footnotesize Total call count & \footnotesize How many individual vocalizations can you detect in this audio? & \footnotesize WABAD \\
  \footnotesize Temporal species ordering & \footnotesize List the species in the order they first vocalize (scientific). & \footnotesize WABAD \\
  \footnotesize Frequency range prediction & \footnotesize What is the overall frequency range of the vocalizations? & \footnotesize WABAD \\
  \footnotesize Per-species frequency ranges & \footnotesize What is the frequency range of each species' vocalizations? & \footnotesize WABAD \\
  \footnotesize Species summary (count + freq.) & \footnotesize Describe each species with call counts and frequency ranges. & \footnotesize WABAD, DCLDE 2026 \\
  \footnotesize AudioSet non-bio multilabel detection & \footnotesize Which of these non-animal sounds are present? \{\{audioset\_option\_choices\}\} & \footnotesize AudioSet \\
  \footnotesize F0 summary statistics & \footnotesize What is the mean fundamental frequency of this vocalization? & \footnotesize F0 Bioacoustic \\
  \footnotesize F0 species-conditioned description & \footnotesize Identify the species and describe the F0 range and duration. & \footnotesize F0 Bioacoustic \\
  \footnotesize Bird presence (binary) & \footnotesize Is there a bird vocalizing in this recording? Yes or No. & \footnotesize Xeno-Canto, Watkins \\
  \footnotesize Marine mammal presence (binary) & \footnotesize Are there whale or dolphin sounds in this recording? Yes or No. & \footnotesize Watkins, Xeno-Canto, Animal Sound Archive \\
  \footnotesize Mammal presence (binary) & \footnotesize Does this recording contain mammal vocalizations? Yes or No. & \footnotesize Watkins, Animal Sound Archive, iNaturalist, Xeno-Canto \\
  \footnotesize Insect presence (binary) & \footnotesize Does this recording contain insect sounds? Yes or No. & \footnotesize Xeno-Canto, iNaturalist, Watkins \\
  \footnotesize Amphibian presence (binary) & \footnotesize Does this recording contain amphibian vocalizations? Yes or No. & \footnotesize Xeno-Canto, iNaturalist, Watkins \\
  \footnotesize Animal presence (binary) & \footnotesize Does this recording contain animal vocalizations? Yes or No. & \footnotesize Xeno-Canto \\
\end{longtable}

\subsection{LLM-based Data Construction} \label{sec:synthetic-construction}
\subsubsection{Overview}\label{llm_overview}
\begin{figure}[h]
    \centering
    \includegraphics[width=0.9\linewidth]{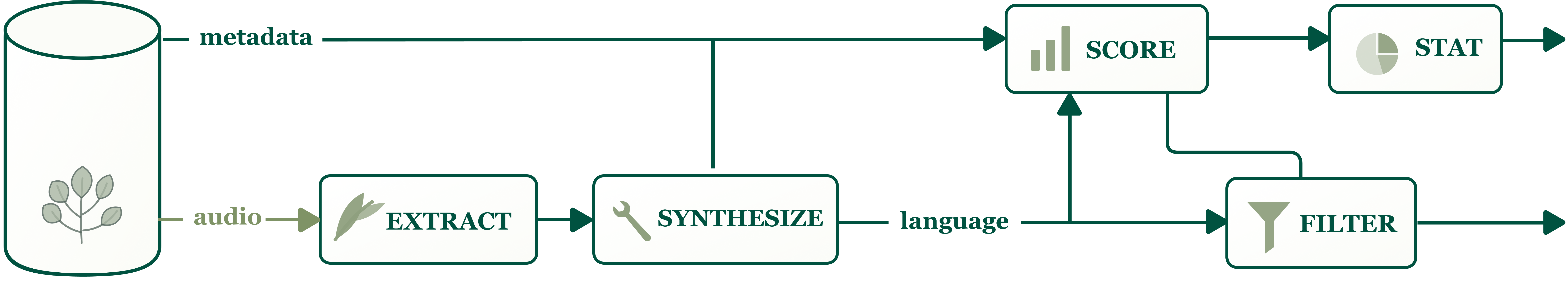}
    \caption{Overview of the LLM-based data synthesis process.}
    \label{fig:pipeline}
\end{figure}

\begin{longtable}{@{}p{3.2cm} p{4.8cm} p{4.5cm}@{}}
\caption{LLM-generated datasets used for training tasks, with a representative prompt.}
\label{tab:llm-datasets-by-task} \\
\toprule
\textbf{Task} & \textbf{Example Prompt} & \textbf{Datasets} \\
\midrule
\endfirsthead
\toprule
\textbf{Task} & \textbf{Example Prompt} & \textbf{Datasets} \\
\midrule
\endhead
\midrule
\multicolumn{3}{r@{}}{\footnotesize\itshape Continued on next page} \\
\endfoot
\bottomrule
\endlastfoot
  \footnotesize SNR category (fixed-bin MCQ) & \footnotesize What is the signal-to-noise ratio (SNR) of the bird call in this recording? Options: Clean, Good, Fair, Poor. & \footnotesize Xeno-Canto, iNaturalist \\
  \footnotesize SNR category (custom-bin MCQ) & \footnotesize How would you classify the SNR of the bird call in this recording? Options use LLM-defined dB ranges. & \footnotesize Xeno-Canto \\
  \footnotesize SNR threshold (binary) & \footnotesize Is the SNR of the bird call in this recording above \{\{threshold\}\} dB? & \footnotesize Xeno-Canto \\
  \footnotesize SNR prediction (open-ended) & \footnotesize Estimate the signal-to-noise ratio (SNR) in dB of the sound event in this recording. & \footnotesize Xeno-Canto \\
  \footnotesize Vocalization description MCQ & \footnotesize Which of the following best describes the vocalization in this recording? Options: \{\{acoustic\_descriptions\}\}. & \footnotesize AnimalSpeak Pseudovox \\
  \footnotesize F0-aware vocalization description MCQ & \footnotesize Which description best matches the sound in this recording? Options include qualitative acoustic character and F0 information. & \footnotesize F0 Bioacoustic \\
  \footnotesize Field-note acoustic description MCQ & \footnotesize Which option most accurately describes the acoustic character of this sound? Options: \{\{acoustic\_descriptions\}\}. & \footnotesize Xeno-Canto, iNaturalist \\
  \footnotesize Behavior / vocalization type MCQ (Hard) & \footnotesize What type of vocalization is the \{\{species\}\} producing in this recording? Options: \{\{behavior\_choices\}\}. & \footnotesize Xeno-Canto, iNaturalist \\
  \footnotesize Semantic captioning & \footnotesize What can you hear in this recording? & \footnotesize Xeno-Canto, iNaturalist \\
  \footnotesize Vocalization location MCQ & \footnotesize At what point in the clip can you hear \{\{species\}\}? Options: near the start, middle, near the end, or absent. & \footnotesize WABAD, Synthetic Strong Detection, Synthetic Detect Diarize \\
  \footnotesize Vocal co-occurrence (binary) & \footnotesize Do \{\{species\_a\}\} and \{\{species\_b\}\} vocalize at the same time in this recording? & \footnotesize WABAD, Synthetic Strong Detection, Synthetic Detect Diarize \\
  \footnotesize Vocal dominance (open-ended) & \footnotesize Which species vocalizes the most in this recording? & \footnotesize WABAD, Synthetic Strong Detection, Synthetic Detect Diarize \\
  \footnotesize Vocal dominance MCQ & \footnotesize Which of these species calls most frequently in this clip? Options: \{\{species\_choices\}\}. & \footnotesize WABAD, Synthetic Strong Detection, Synthetic Detect Diarize \\
  \footnotesize Species by vocalization order (open-ended) & \footnotesize Which species produces the \{\{ordinal\}\} vocalization in this recording? & \footnotesize WABAD, Synthetic Strong Detection, Synthetic Detect Diarize \\
  \footnotesize Species by vocalization order MCQ & \footnotesize Which of the following species calls \{\{ordinal\}\} in this clip? Options: \{\{species\_choices\}\}. & \footnotesize WABAD, Synthetic Strong Detection, Synthetic Detect Diarize \\
  \footnotesize Highest-pitched species (open-ended) & \footnotesize Which species produces the highest-pitched vocalization in this recording? & \footnotesize WABAD \\
  \footnotesize Highest-pitched species MCQ & \footnotesize Which of these animals produces the highest-pitched vocalization in this clip? Options: \{\{species\_choices\}\}. & \footnotesize WABAD \\
  \footnotesize Lowest-pitched species (open-ended) & \footnotesize Which species produces the lowest-pitched vocalization in this recording? & \footnotesize WABAD \\
  \footnotesize Lowest-pitched species MCQ & \footnotesize Which of these animals produces the lowest-pitched vocalization in this clip? Options: \{\{species\_choices\}\}. & \footnotesize WABAD \\
  \footnotesize Longest vocalization species (open-ended) & \footnotesize Which species produces the longest vocalization in this recording? & \footnotesize WABAD, Synthetic Strong Detection, Synthetic Detect Diarize \\
  \footnotesize Longest vocalization species MCQ & \footnotesize Which of these species makes the longest call in this clip? Options: \{\{species\_choices\}\}. & \footnotesize WABAD, Synthetic Strong Detection, Synthetic Detect Diarize \\
  \footnotesize Structural captioning & \footnotesize Describe the temporal structure and event organization in this recording. & \footnotesize WABAD, Synthetic Strong Detection, Synthetic Detect Diarize \\
\end{longtable}

\subsection{Synthetic Audio Creation} \label{sec:synthetic-audio}
Synthetic scenes were constructed by mixing extracted vocalizations (``pseudovox'') into real environmental backgrounds. Pseudovox were obtained by source-separating Xeno-canto recordings using BirdMixIT \citep{denton2022improving} and extracting short activity windows exactly as proposed in DRASDIC \citep{drasdic}; we used a filtered subset of these clips, each tagged with a species label (predicted by sl-efficientnet-bio \citep{miron2026avex} and kept where it matched the metadata), an acoustic cluster ID reused from DRASDIC, and a duration. Background recordings were drawn from SilentCities. To construct a scene, we sample a background, optionally mix two backgrounds for acoustic diversity, and insert a configurable number of pseudovox events at target SNRs sampled from a Gaussian mixture (range $-20$~dB to $+15$~dB), with random resampling, optional time reversal, and randomized inter-event gaps. All audio is rendered at 16~kHz.
Four datasets were generated from this pipeline. \textbf{Call type MCQ} and \textbf{Binary call type detection} present the model with one or more example pseudovox clips of a target call type and ask it to identify or detect that call type in a separate audio clip. \textbf{Same/different call type} presents two pseudovox clips and asks whether they share a call type. For these three tasks, ``call type'' is operationalized as the intersection of species and acoustic cluster (``fine mode''); negatives are sampled from unrelated species--cluster combinations. \textbf{Few-shot detection} is a multi-label variant: the model is given example clips for 1--4 candidate sounds plus a query scene of 3--10~s and must identify which (if any) of the candidates appear in the scene; queries optionally include a short clip of the scene's background environment to help the model distinguish target events from ambient noise. All synthetic datasets are split-safe: pseudovox come from the \texttt{train\_\allowbreak unseen} split of Xeno-canto and are not used in evaluation.

\section{Training}
\subsection{Training Hyperparameters} \label{sec:hypers}

The stage-1 LALM was trained on only open-ended species classification data, with a global batch size of 128 for 350k steps, at a peak lr of $1e-4$ with 25,000 steps of warmup and cosine decay to a minimum lr of $3e-5$. The stage-2 LALM was trained for 1.5 million steps and a global batch size of 128 at a peak lr of $1e-4$ with 25,000 steps of warmup and cosine decay to a minimum lr of $3e-5$. Audio was limited to 10-second clips. This completed stage-2 formed the base model for fine-tuning.

\subsection{Fine-tuning Hyperparameters} \label{sec:ft_hypers}
The per-tier fine-tunings were instantiated from the stage 2 checkpoint for 2 epochs, using a global batch size of 128. We used an initial learning rate of $1e^{-6}$, 15,000 warmup steps to a peak learning rate of $1e^{-4}$, and cosine decay to a minimum learning rate of $1e^{-5}$. LoRA fine-tuning was applied to the query, key, value, and output projection layers, using rank 32, $\alpha=32$, and dropout 0.1. Audio was limited to 30-second clips.

\subsection{Augmentations}
We apply audio augmentations during stage-2 training to improve robustness to background noise and overlapping vocalizations. With probability 0.4, we add background noise sampled from a pool of environmental and general-audio noise corpora, including DEMAND, IDMT, TUT, UrbanSound, OrCA/ship-noise datasets, WHAM, AudioSet, and SilentCities, at an SNR sampled uniformly from $[-10, 25]$ dB. Noise datasets were curated in advance, including SilentCities being processed with sl-efficient-all model \citep{miron2026avex} to filter clips with animal vocalizations. For detection-style tasks, we additionally apply a noise-masking variant with probability 0.15, replacing the target signal with background noise and training the model to output the corresponding negative label. We also use task-aware mixup with probability 0.3, mixing up to three compatible partner recordings into a primary example with Beta-distributed weights ($\alpha=0.4$). For open-ended detection tasks, partner labels are unioned into the target answer; for option-based detection tasks, mixed-in labels are added only when they appear among the provided candidate options, otherwise they serve as distractor background. Noise augmentation is applied across all training stages, while MixUp is only applied in stage-2 onward and only for detection tasks.

\subsection{Computational resources}
For each fine-tune, 8 NVIDIA H100 80GB GPUs were used. For Tier 1 and Tier 4, the fine-tuning time was approximately 1 day. For Tier 2 and Tier 3, the fine-tuning time was approximately 3 days. Thus, the total train time was approximately 64 H100 days.

\subsection{Corpus statistics}\label{sec:corpus}
\begin{figure}
    \centering
    \includegraphics{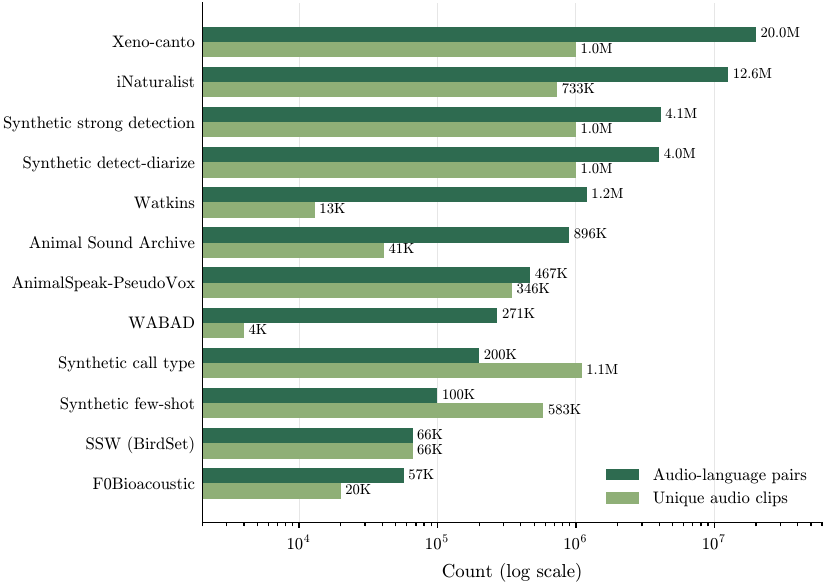}
    \caption{\dataset{} data sources, with the number of audio-language pairs and unique audio clips contributed by each (log scale).}
    \label{fig:corpus_size}
\end{figure}

Figure \ref{fig:corpus_size} displays the audio-language and unique audio counts from each data source.

\begin{figure}
    \centering
    \includegraphics{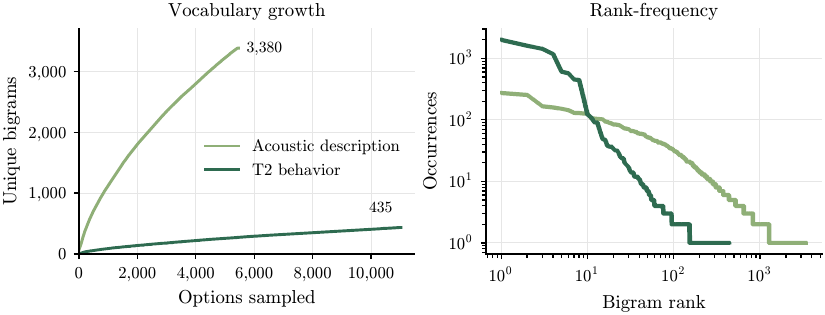}
\caption{Lexical diversity of LLM-synthesized acoustic-description and behavior MCQ options. \textbf{Left:} number of unique word bigrams as more options are sampled. \textbf{Right:} bigram rank--frequency distribution (log--log). Acoustic descriptions keep introducing new bigrams and have a long, flat tail, whereas behavior options reuse a smaller vocabulary.}
\label{fig:div}
\end{figure}

Even though we create audio-language pairs from several datasets that have been used before to train bioacoustic audio-language models, our LLM-based text construction pipeline allows us to leverage more information, including more nuanced and detailed behavioral and acoustic descriptions. Figure~\ref{fig:div} shows the 2-gram diversity for two of the tasks that leveraged the LLM-based synthesis: acoustic and behavioral description.

\section{\benchmark{} Construction } \label{sec:benchmark-creation}

\subsection{Data Curation}
\paragraph{Tier 1 benchmark processing.}
For expert acoustic description matching, we constructed four-way multiple-choice questions from published carrion crow vocal repertoire descriptions. Each example presents one query recording and four candidate expert descriptions from the same species, with one correct answer. Descriptions were sourced from the original study~\citep{carrion_crows}, with small modifications: we condensed the caws caw0--5, grunt0--5Q, and grunt0--5L into one group each for evaluation, to avoid asking the model to distinguish calls primarily by duration. We changed descriptions for the few calls that referenced another call to be standalone descriptions, and otherwise left descriptions as given by the authors.

For mean F0 prediction, we used the F0 annotation dataset by~\citet{Marxer}. We parsed each F0 contour, removed invalid values, computed the mean frequency over valid contour points, and rounded the target to the nearest 10 Hz. We report separate splits for taxa represented in training and a held-out split containing a single species, the spotted hyena (\textit{Crocuta crocuta}). This split is specific to the $f_0$ task and separate from the species-identity held-out taxa in Appendix~\ref{sec:held-out}.

For SNR prediction, we use the LLM-based pipeline to mine SNR annotations that reference the dominant vocalization in the recording from the field notes in Xeno-canto. These follow ISO 226, and Xeno-canto provides a guide for their computation, making them likely consistent.\footnote{\url{https://xeno-canto.org/article/275}} We create an open-ended regression SNR evaluation, as well as an MCQ-based evaluation, that provides 4 bins with indicative SNR ranges as options.

For acoustic description, we use the LLM-based pipeline to mine acoustic descriptions from Xeno-canto sample field notes, and reformat them into captions or MCQs. The synthesis LLM retains only directly acoustic information such as pitch, texture, frequency movement, and rhythm, while excluding species names, call-type labels, behavioral context, and other semantic cues.

\paragraph{Tier 2 benchmark processing.}
For taxon presence detection, we used held-out Xeno-canto validation recordings and constructed balanced binary yes/no tasks for bird, mammal, insect, and amphibian vocalizations. Positives contain the target taxon according to focal label; negatives are sampled from other taxa in the held-out pool, with additional filtering on background species and field notes in the metadata.

For call-type evaluation, we derived binary alarm-call, flight-call, and begging-call presence tasks, plus a fixed-vocabulary multi-label task over \texttt{alarm call}, \texttt{flight call}, \texttt{begging call}, \texttt{song}, and \texttt{call}. These examples are based on BEANS-Zero call-type variants linked to Xeno-canto metadata.

We additionally include two LLM-synthesized Tier 2 semantic evaluation splits from Xeno-canto and iNaturalist BEANS-Zero/new held-out pools. Semantic captioning asks for natural-language descriptions of audible biological content, including call type, behavioral context, acoustic character, life stage when acoustically relevant, and background species when supported by the metadata. Behavior MCQs ask which audible behavior or context best matches the recording, using plausible distractors generated from the same metadata regime. These tasks differ from Tier 1 acoustic captions because semantic labels and behavioral meaning are allowed when supported by \texttt{fieldNotes}, \texttt{occurrenceRemarks}, and species metadata. We removed generations judged insufficiently audible, interesting, or difficult using an LLM-judge filter.

\paragraph{Tier 3 benchmark processing.}
For single-audio scene understanding, we used recordings from Birdeep, Powdermill, NocturnalBirdMigration, and BirdVox full-night, cropped randomly to between 1 and 30 seconds. Their ground-truth selection tables provide species labels, temporal extents, and, for Birdeep, frequency ranges; synthesis prompts convert these annotations into natural questions while excluding ambiguous rows such as ties, boundary-clipped events that could change the answer, or cases with insufficient annotation support.

Tier 3 includes LLM-based text synthesis for species counting, total vocalization counting, per-species vocalization counting, temporal referring (start, middle, or end of recording), vocalization co-occurrence, species by vocalization order (e.g. which species vocalizes third), species by highest pitch, species by lowest pitch, species by longest vocalization, species by vocalization frequency or dominance, and structural captioning, which provides an overview of vocalizations in the recording. We include both open-ended and MCQ variants where appropriate, the latter of which provide reference species as options, with a random number of them being other species that are present in the recording. We also create tasks with templates, namely listing all species in the recording, listing the frequency range, and listing the species, frequency range, and vocalization count for each species.

\paragraph{Tier 4 benchmark processing.}
For gibbon few-shot detection, we used Hainan gibbon recordings and held out fixed support examples for three call categories: multiple-pulse calls, single-pulse calls, and duets. Each query asks which support sound, if any, appears in the target recording. We include both the full naturally imbalanced split and a balanced split for headline evaluation.

For DCASE few-shot detection, we used DCASE 2021 Task 5 sound-event examples. Each row presents audio exemplars for candidate sound classes, optionally a background clip, and a query recording. The model predicts a set of option labels or \texttt{None}. The released split is balanced between present and none examples.

For four-way call-type matching, we constructed audio-exemplar multiple-choice tasks for giant otters and carrion crows. Each example provides four labeled support clips, one per candidate call type, and a query clip from the correct class. Answer labels are balanced across A--D where possible.

For unseen-species matching, we used held-out Xeno-canto and iNaturalist taxa not included in the training mixture, based on same unseen taxa held out in NatureLM-audio \citep{robinsonnaturelm}. Each example presents four species exemplars and a query recording; the model must select the matching species. Distractors are sampled from other held-out species, prioritizing hard, related-species negatives when available to make a more challenging task.

\subsubsection{Evaluation tasks and abbreviations}\label{sec:evals}
\begingroup
\footnotesize
\setlength{\tabcolsep}{4pt}
\begin{longtable}{@{}>{\raggedright\arraybackslash}p{0.33\linewidth}>{\raggedright\arraybackslash}p{0.27\linewidth}>{\raggedright\arraybackslash}p{0.13\linewidth}r >{\raggedright\arraybackslash}p{0.10\linewidth}@{}}
\caption{BEANS-Next evaluation task labels used in result tables. Labels include the higher-level table group followed by the abbreviated column name used in the main results.}\\
\toprule
Table label & Task & Format & Samples & Metric \\
\midrule
\endfirsthead
\toprule
Table label & Task & Format & Samples & Metric \\
\midrule
\endhead
\bottomrule
\endfoot

\multicolumn{5}{@{}l}{\textit{Tier 1: Acoustic}}\\
\texttt{Acoustic description / Caption} & Acoustic caption generation & Caption / OE & 213 & CIDEr \\
\texttt{Acoustic description / Desc.\ MCQ} & Field-note acoustic description matching & MCQ & 290 & Accuracy \\
\texttt{Acoustic description / Crow} & Carrion crow expert description matching & MCQ & 200 & Accuracy \\
\texttt{SNR / MCQ} & SNR category prediction & MCQ & 607 & Accuracy \\
\texttt{SNR / MAE (dB)} & SNR estimation & OE / regression & 397 & MAE (dB) \\
Mean $f_0$ / \texttt{Seen} & Mean $f_0$ regression for seen taxa & OE / regression & 2,086 & MAE (Hz) \\
Mean $f_0$ / \texttt{Held-out} & Mean $f_0$ regression for the held-out species & OE / regression & 571 & MAE (Hz) \\

\addlinespace
\multicolumn{5}{@{}l}{\textit{Tier 2: Biological recognition}}\\
\texttt{Taxon presence / Bird} & Bird presence & Binary & 2,000 & Accuracy \\
\texttt{Taxon presence / Mamm.} & Mammal presence & Binary & 448 & Accuracy \\
\texttt{Taxon presence / Amph.} & Amphibian presence & Binary & 1,856 & Accuracy \\
\texttt{Taxon presence / Insect} & Insect presence & Binary & 636 & Accuracy \\
\texttt{Call-type presence / Alarm} & Alarm-call presence & Binary & 974 & Accuracy \\
\texttt{Call-type presence / Flight} & Flight-call presence & Binary & 2,000 & Accuracy \\
\texttt{Call-type presence / Beg.} & Begging-call presence & Binary & 586 & Accuracy \\
\texttt{Call type} & Fixed-vocabulary call-type prediction & Multi-label & 1,000 & Macro-F1 \\
\texttt{Behav.} & Behavior or call-type recognition & MCQ & 1,169 & Accuracy \\
\texttt{Caption} & Semantic caption generation & Caption / OE & 3,000 & CIDEr \\

\addlinespace
\multicolumn{5}{@{}l}{\textit{Tier 3: Counting, temporal, and scene/open}}\\
\texttt{Counting / \#S} & Number of species present & OE / numeric & 1,000 & MAE \\
\texttt{Counting / \#V} & Total number of vocalizations & OE / numeric & 993 & MAE \\
\texttt{Counting / \#V/S} & Number of vocalizations per species & Open table & 866 & MAE \\
\texttt{Temporal / Referring} & Identify a vocalization from a temporal reference & MCQ & 893 & Accuracy \\
\texttt{Temporal / Overlap} & Determine whether vocalizations overlap in time & Binary & 883 & Accuracy \\
\texttt{Scene / open / Range} & Frequency range of each species in the scene & Open list & 855 & Species-band IoU \\
\texttt{Scene / open / Sum.} & Ordered summary of species activity & Caption / OE & 1,000 & CIDEr \\
\texttt{Scene / open / Caption} & Structural scene caption generation & Caption / OE & 1,000 & CIDEr \\

\addlinespace
\multicolumn{5}{@{}l}{\textit{Tier 3: Species ID by characteristic}}\\
\texttt{Species ID / OE / Order} & Identify species by order of vocalization & OE & 942 & Accuracy \\
\texttt{Species ID / OE / High} & Identify the highest-pitched species & OE & 868 & Accuracy \\
\texttt{Species ID / OE / Low} & Identify the lowest-pitched species & OE & 803 & Accuracy \\
\texttt{Species ID / OE / Long} & Identify the longest-vocalizing species & OE & 829 & Accuracy \\
\texttt{Species ID / OE / Freq.} & Identify the most vocally active species & OE & 844 & Accuracy \\
\texttt{Species ID / MCQ / Order} & Identify species by order of vocalization & MCQ & 832 & Accuracy \\
\texttt{Species ID / MCQ / High} & Identify the highest-pitched species & MCQ & 782 & Accuracy \\
\texttt{Species ID / MCQ / Low} & Identify the lowest-pitched species & MCQ & 594 & Accuracy \\
\texttt{Species ID / MCQ / Long} & Identify the longest-vocalizing species & MCQ & 863 & Accuracy \\
\texttt{Species ID / MCQ / Freq.} & Identify the most vocally active species & MCQ & 873 & Accuracy \\

\addlinespace
\multicolumn{5}{@{}l}{\textit{Tier 4: Multi-audio few-shot}}\\
\texttt{Few-shot task / Gibbons} & Gibbon few-shot detection & Few-shot multi-label & 200 & Macro-F1 \\
\texttt{Few-shot task / Otters} & Giant otter call-type recognition & Multi-audio MCQ & 200 & Accuracy \\
\texttt{Few-shot task / DCASE} & DCASE few-shot detection & Few-shot multi-label & 200 & Macro-F1 \\
\texttt{Few-shot task / Crows} & Carrion crow call-type recognition & Multi-audio MCQ & 200 & Accuracy \\
\texttt{Few-shot task / Unseen} & Unseen species identification & Multi-audio MCQ & 200 & Accuracy \\

\addlinespace
\multicolumn{5}{@{}l}{\textit{Additional evals}}\\
\texttt{CBI} & BEANS-Zero Cornell Bird Identification & OE classification & 3,620 & Accuracy \\
\texttt{HBDB} & BEANS-Zero HumBugDB mosquito classification & OE classification & 1,859 & Accuracy \\
\texttt{ESC50} & BEANS-Zero ESC-50 environmental sound classification & OE classification & 400 & Accuracy \\
\texttt{Call-type} & BEANS-Zero bird call-vs-song classification & OE classification & 1,000 & Accuracy \\
\texttt{Lifestage} & BEANS-Zero bird life-stage classification & OE classification & 466 & Accuracy \\
\texttt{N-indv} & BEANS-Zero zebra-finch individual-count classification & OE classification & 1,160 & Accuracy \\
\texttt{BirdSet} & BirdSet classification & OE classification & 117,092 & Accuracy \\
\end{longtable}
\endgroup

\begin{figure}[t]
  \centering
    \centering
    \includegraphics{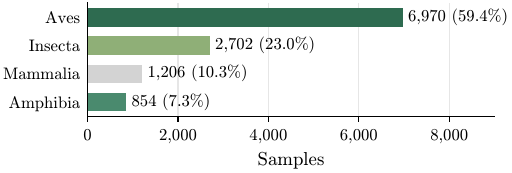}
    \caption{Taxonomic class distribution of \benchmark{} samples with full taxonomic metadata (Tiers 2 and 4).}
    \label{fig:taxonomy_pie}
  \label{fig:taxonomy}
\end{figure}

\begin{figure}[t]
  \centering
  \includegraphics{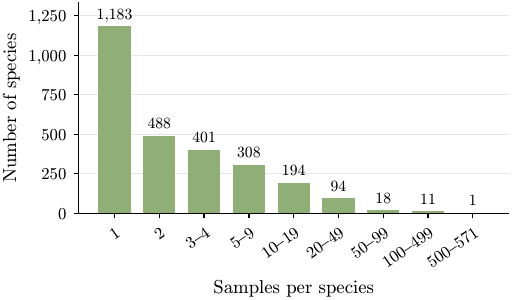}
  \caption{
    Species frequency distribution across \textsc{BEANS-Next}: number of species by samples per species. Of the 2{,}698 unique species, 1{,}183 (44\%) appear only once, while a few focal species (e.g., \textit{Crocuta crocuta}, \textit{Corvus corone}) contribute hundreds of samples, stressing few-shot and zero-shot generalization rather than memorization of frequent taxa.
  }
  \label{fig:species_longtail}
\end{figure}

\begin{figure}[t]
  \centering
  \includegraphics{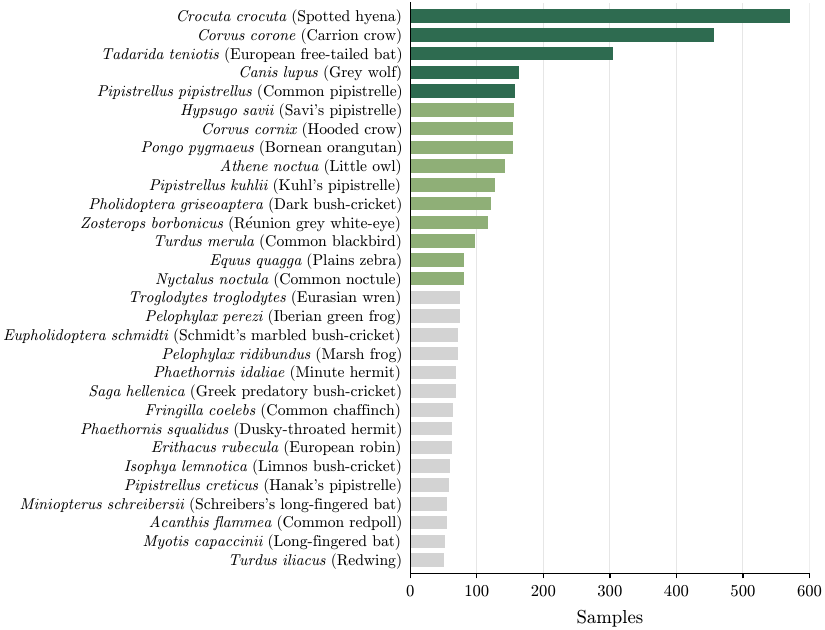}
  \caption{
    Top 30 species by sample count in \textsc{BEANS-Next}.
    High-count species are predominantly focal study subjects from
    controlled recording campaigns (e.g., \textit{Corvus corone} from
    the crow call-type dataset, \textit{Tadarida teniotis} and \textit{Pipistrellus} spp.\ from bat monitoring).
    The rapid drop-off after the top few species illustrates the
    long-tail character of the overall species distribution.
  }
  \label{fig:top_species}
\end{figure}

\definecolor{pinegreen}{HTML}{1F4D3A}
\definecolor{sagegreen}{HTML}{9CAF88}
\definecolor{lightgraygreen}{HTML}{F3F4F1}

\newcommand{\yes}{\textbf{Yes}}
\newcommand{\no}{\textcolor{gray}{No}}

\begin{table}[t]
\centering
\small
\setlength{\tabcolsep}{5pt}
\renewcommand{\arraystretch}{1.15}

\begin{tabular}{lcccc}
\toprule
\rowcolor{pinegreen}
\textcolor{white}{\textbf{Capability}} &
\textcolor{white}{\textbf{BEANS}} &
\textcolor{white}{\textbf{BirdSet}} &
\textcolor{white}{\textbf{WoW-Bench}} &
\textcolor{white}{\textbf{BEANS-Next}} \\
\midrule

Species recognition
& \cellcolor{sagegreen!25}\yes
& \cellcolor{sagegreen!25}\yes
& \no
& \cellcolor{sagegreen!40}\yes \\

Acoustic measurement
& \no
& \no
& \cellcolor{sagegreen!25}\yes
& \cellcolor{sagegreen!40}\yes \\

Temporal reasoning
& \no
& \no
& \no
& \cellcolor{sagegreen!40}\yes \\

Relational reasoning
& \no
& \no
& \no
& \cellcolor{sagegreen!40}\yes \\

Multi-audio reasoning
& \no
& \no
& \no
& \cellcolor{sagegreen!40}\yes \\

In-context learning
& \no
& \no
& \no
& \cellcolor{sagegreen!40}\yes \\

\bottomrule
\end{tabular}

\caption{
Comparison of benchmark coverage across bioacoustic audio-language capabilities. Existing benchmarks primarily evaluate species recognition or isolated acoustic perception, while BEANS-Next evaluates a broader capability space required by ethological workflows.
}

\label{tab:benchmark_capability_coverage}
\end{table}

\subsection{List of \benchmark{} held-out taxa}\label{sec:held-out}

\headinggap
\noindent\textbf{Held-out species:}
\begin{multicols}{3}
\footnotesize
\begin{enumerate}[leftmargin=*, itemsep=0pt, parsep=0pt, topsep=2pt]
\item \textit{Aethopyga shelleyi}
\item \textit{Arachnothera dilutior}
\item \textit{Sitta castanea}
\item \textit{Carpodacus rodopeplus}
\item \textit{Aethopyga ignicauda}
\item \textit{Pachycephala soror}
\item \textit{Herpsilochmus roraimae}
\item \textit{Amazona dufresniana}
\item \textit{Metallura aeneocauda}
\item \textit{Thlypopsis fulviceps}
\item \textit{Monarcha frater}
\item \textit{Kleinothraupis reyi}
\item \textit{Aplonis magna}
\item \textit{Phylloscopus misoriensis}
\item \textit{Agapornis pullarius}
\item \textit{Amazona versicolor}
\item \textit{Saltator cinctus}
\item \textit{Xiphocolaptes falcirostris}
\item \textit{Passer insularis}
\item \textit{Chalcomitra balfouri}
\item \textit{Arremonops tocuyensis}
\item \textit{Atlapetes meridae}
\item \textit{Colluricincla obscura}
\item \textit{Saltator maxillosus}
\item \textit{Philemon meyeri}
\item \textit{Thamnophilus insignis}
\item \textit{Aulacorhynchus whitelianus}
\item \textit{Sirystes subcanescens}
\item \textit{Sporophila nigrorufa}
\item \textit{Zoothera mollissima}
\item \textit{Thlypopsis inornata}
\item \textit{Picumnus spilogaster}
\item \textit{Columba arquatrix}
\item \textit{Petrochelidon rufocollaris}
\item \textit{Pyrrhura griseipectus}
\item \textit{Myiothlypis chrysogaster}
\item \textit{Thripophaga amacurensis}
\item \textit{Herpsilochmus motacilloides}
\item \textit{Progne dominicensis}
\item \textit{Heliodoxa branickii}
\item \textit{Asthenes arequipae}
\item \textit{Gerygone fusca}
\item \textit{Otus thilohoffmanni}
\item \textit{Inezia subflava}
\item \textit{Charadrius montanus}
\item \textit{Petroica polymorpha}
\item \textit{Symposiachrus vidua}
\item \textit{Dicrurus lophorinus}
\item \textit{Pycnonotus penicillatus}
\item \textit{Melanerpes herminieri}
\item \textit{Zosterops mysorensis}
\item \textit{Oenanthe xanthoprymna}
\item \textit{Artamus monachus}
\item \textit{Caprimulgus pulchellus}
\item \textit{Psarocolius cassini}
\item \textit{Symposiachrus infelix}
\item \textit{Zosterops cinereus}
\item \textit{Circus cinereus}
\item \textit{Geotrygon chrysia}
\item \textit{Microspingus trifasciatus}
\item \textit{Pternistis harwoodi}
\item \textit{Ceblepyris caesius}
\item \textit{Ficedula disposita}
\item \textit{Treron affinis}
\item \textit{Geokichla wardii}
\item \textit{Campethera bennettii}
\item \textit{Alcedo semitorquata}
\item \textit{Buteo japonicus}
\item \textit{Apus bradfieldi}
\item \textit{Pterocles personatus}
\item \textit{Melaniparus fringillinus}
\item \textit{Poecile hypermelaenus}
\item \textit{Circus buffoni}
\item \textit{Pycnonotus blanfordi}
\item \textit{Machlolophus aplonotus}
\item \textit{Estrilda ochrogaster}
\item \textit{Touit batavicus}
\item \textit{Mirafra gilletti}
\item \textit{Pternistis icterorhynchus}
\item \textit{Accipiter collaris}
\item \textit{Knipolegus lophotes}
\item \textit{Nothoprocta taczanowskii}
\item \textit{Pachycephala modesta}
\item \textit{Vanellus tricolor}
\item \textit{Caprimulgus andamanicus}
\item \textit{Ardenna grisea}
\item \textit{Mixornis kelleyi}
\item \textit{Cinnyris johannae}
\item \textit{Recurvirostra novaehollandiae}
\item \textit{Sitta leucopsis}
\item \textit{Petroica pusilla}
\item \textit{Amazilia luciae}
\item \textit{Melaniparus fasciiventer}
\item \textit{Egretta picata}
\item \textit{Columba pollenii}
\item \textit{Rallus madagascariensis}
\item \textit{Heliodoxa gularis}
\item \textit{Carpodacus roseus}
\item \textit{Zosterops chloronothos}
\item \textit{Pachycephala lorentzi}
\item \textit{Saucerottia cyanura}
\item \textit{Cinclosoma marginatum}
\item \textit{Bucco noanamae}
\item \textit{Certhia nipalensis}
\item \textit{Pachycephala lanioides}
\item \textit{Carpodacus trifasciatus}
\item \textit{Chorthippus acroleucus}
\item \textit{Chlidonias albostriatus}
\item \textit{Hirundo domicola}
\item \textit{Falco concolor}
\item \textit{Dryocopus schulzii}
\item \textit{Rhyticeros undulatus}
\item \textit{Quiscalus nicaraguensis}
\item \textit{Cisticola brunnescens}
\item \textit{Knipolegus cyanirostris}
\item \textit{Ardenna carneipes}
\item \textit{Lybius rubrifacies}
\item \textit{Climacteris melanurus}
\item \textit{Puffinus opisthomelas}
\item \textit{Manorina melanotis}
\item \textit{Celebesica abbotti}
\item \textit{Otus mayottensis}
\item \textit{Trachyphonus margaritatus}
\item \textit{Oenanthe dubia}
\item \textit{Chloropsis flavipennis}
\item \textit{Ploceus alienus}
\item \textit{Phalacrocorax varius}
\item \textit{Ploceus pelzelni}
\item \textit{Merops mentalis}
\item \textit{Passer gongonensis}
\item \textit{Myzomela cineracea}
\item \textit{Pachycephala feminina}
\item \textit{Brachypteryx sinensis}
\item \textit{Lonchura flaviprymna}
\item \textit{Ninox natalis}
\item \textit{Myrmelastes caurensis}
\item \textit{Buteo trizonatus}
\item \textit{Apalis chariessa}
\item \textit{Ficedula nigrorufa}
\item \textit{Pica mauritanica}
\item \textit{Anthreptes reichenowi}
\item \textit{Sholicola major}
\item \textit{Vireo osburni}
\item \textit{Anas capensis}
\item \textit{Ducula luctuosa}
\item \textit{Lanius newtoni}
\item \textit{Odontophorus dialeucos}
\item \textit{Bostrychia olivacea}
\item \textit{Cinnyris tsavoensis}
\item \textit{Ploceus heuglini}
\item \textit{Myzomela nigrita}
\item \textit{Falco cherrug}
\item \textit{Ixobrychus sturmii}
\item \textit{Rhipidura semirubra}
\item \textit{Haematopus chathamensis}
\item \textit{Anthus brachyurus}
\item \textit{Oenanthe lugens}
\item \textit{Columba rupestris}
\item \textit{Rhyticeros subruficollis}
\item \textit{Zosterops vellalavella}
\item \textit{Anthus sokokensis}
\item \textit{Phaethornis idaliae}
\item \textit{Picus dedemi}
\item \textit{Muscicapa segregata}
\item \textit{Cyanomitra bannermani}
\item \textit{Polioptila facilis}
\item \textit{Platysteira albifrons}
\item \textit{Dicaeum pygmaeum}
\item \textit{Puffinus assimilis}
\item \textit{Rhipidura kubaryi}
\item \textit{Ploceus katangae}
\item \textit{Canis lupaster}
\item \textit{Hyla andersonii}
\item \textit{Ranoidea nudidigita}
\item \textit{Ranoidea aurea}
\item \textit{Litoria tyleri}
\item \textit{Dendropsophus joannae}
\item \textit{Okanagana occidentalis}
\item \textit{Litoria latopalmata}
\item \textit{Magicicada tredecassini}
\item \textit{Orchelimum silvaticum}
\item \textit{Oecanthus celerinictus}
\item \textit{Empidonomus aurantioatrocristatus}
\item \textit{Bufotes boulengeri}
\item \textit{Oecanthus nigricornis}
\item \textit{Myrmothera fulviventris}
\item \textit{Psaltoda adonis}
\item \textit{Rana dalmatina}
\item \textit{Dendropsophus sanborni}
\item \textit{Hyperolius stictus}
\item \textit{Hyperolius pictus}
\item \textit{Hyla eximia}
\item \textit{Leptodactylus natalensis}
\item \textit{Oecanthus californicus}
\item \textit{Hyperolius parallelus}
\item \textit{Gryllus cohni}
\item \textit{Physeter macrocephalus}
\item \textit{Eleutherodactylus unicolor}
\item \textit{Gryllus bermudensis}
\item \textit{Anas penelope}
\end{enumerate}
\end{multicols}

\headinggap
\noindent\textbf{Held-out genera:}
\begin{multicols}{3}\footnotesize\begin{enumerate}[leftmargin=*, itemsep=0pt, parsep=0pt, topsep=2pt]\item \textit{Aglaeactis}\item \textit{Drepanorhynchus}\item \textit{Lesbia}\item \textit{Nemobius}\item \textit{Meconema}\item \textit{Pseudochorthippus}\item \textit{Caliechthrus}\item \textit{Pachycare}\item \textit{Rhodothraupis}\item \textit{Astrapia}\item \textit{Probosciger}\item \textit{Amazonetta}\item \textit{Ocyalus}\item \textit{Nandayus}\item \textit{Rhinocrypta}\item \textit{Heterocercus}\item \textit{Jacamaralcyon}\item \textit{Hymenops}\item \textit{Doliornis}\item \textit{Eugerygone}\item \textit{Cryptosylvicola}\item \textit{Taeniopygia}\item \textit{Catharopeza}\item \textit{Eurostopodus}\item \textit{Tylas}\item \textit{Vini}\item \textit{Ptychoramphus}\item \textit{Speculanas}\item \textit{Aphelocephala}\item \textit{Stipiturus}\item \textit{Procarduelis}\item \textit{Rhopophilus}\item \textit{Neopsephotus}\item \textit{Enodes}\item \textit{Leucocarbo}\item \textit{Gymnophaps}\item \textit{Goldmania}\item \textit{Oreomystis}\item \textit{Rhodostethia}\item \textit{Falcipennis}\item \textit{Pachycoccyx}\item \textit{Cryptotympana}\item \textit{Tympanistalna}\item \textit{Cyrtoxipha}\item \textit{Afrixalus}\item \textit{Uperoleia}\item \textit{Urocitellus}\item \textit{Chalcorana}\item \textit{Aiolopus}\item \textit{Speothos}\end{enumerate}\end{multicols}

\headinggap
\noindent\textbf{Held-out families:}
\begin{multicols}{2}
\footnotesize
\begin{enumerate}[leftmargin=*, itemsep=0pt, parsep=0pt, topsep=2pt]
\item \textit{Elachuridae}
\item \textit{Calyptophilidae}
\item \textit{Pelecanoididae}
\item \textit{Phocoenidae}
\item \textit{Alytidae}
\item \textit{Castoridae}
\item \textit{Dicroglossidae}
\item \textit{Suidae}
\item \textit{Prophalangopsidae}
\item \textit{Octodontidae}
\end{enumerate}
\end{multicols}

\section{Evaluation Pipeline and Metrics}
\label{sec:evaluation}

\subsection{Models}\label{apd:models}

\headinggap
\noindent\textbf{NatureLM-audio v1.0.}
NatureLM-audio v1.0~\citep{robinsonnaturelm} is an audio-language foundation model specifically designed for bioacoustics, combining a BEATs audio encoder with a Llama 3.1 8b instruction-tuned language model backbone. The model was trained on paired audio-text data spanning animal vocalizations, speech, and music, and supports a broad range of zero-shot bioacoustic tasks, including species classification, vocalization detection, call-type recognition, and audio captioning. NatureLM-audio was introduced alongside the BEANS-Zero benchmark and demonstrated strong transfer performance on unseen taxa and tasks.

\headinggap
\noindent\textbf{NatureLM-audio v1.1.}
We train NatureLM-audio v1.1 for the first time in this work. It retains the architecture of NatureLM-audio v1.0 but incorporates infrastructure improvements supporting longer audio and multiple audio files as input, together with newly ingested training data from the same underlying bioacoustic archives. NatureLM-audio v1.1 serves as the common base checkpoint for the tier-specific ROOTS experiments. Project resources and model weights are publicly available through the associated web page.\footnote{\url{https://earthspecies.github.io/naturelm-audio-demo}}

\headinggap
\noindent\textbf{Audio Flamingo Next (AF-Next).}
Audio Flamingo Next~\citep{ghosh_audio_2026} is a large open audio-language model for unified understanding of speech, environmental sounds, and music. AF-Next extends the Audio Flamingo family with substantially larger-scale training data, long-context audio processing (up to 30 minutes), and temporal chain-of-thought reasoning mechanisms grounded in timestamps. The model achieves state-of-the-art or competitive performance across a wide range of audio understanding and reasoning benchmarks. AF-Next builds upon earlier Audio Flamingo models that introduced few-shot and conversational audio-language reasoning capabilities for open-ended audio understanding tasks.
Model checkpoints and project resources are available through NVIDIA research releases and Hugging Face.\footnote{\url{https://huggingface.co/nvidia/audio-flamingo-next-hf}}

\headinggap
\noindent\textbf{Qwen3-Omni Instruct.}
Qwen3-Omni~\citep{xu2025qwen3} is a unified multimodal foundation model supporting text, image, audio, and video understanding and generation within a single architecture. The model employs a ``Thinker--Talker'' mixture-of-experts design intended to jointly support multimodal reasoning and streaming speech generation. The audio subsystem achieves strong performance across audio and audio-visual benchmarks, including competitive results against large proprietary systems.
In our experiments, we use the instruction-tuned Qwen3-Omni variant for zero-shot audio understanding and captioning tasks. Public model implementations and checkpoints are available through the Qwen repositories.\footnote{\url{https://github.com/QwenLM/Qwen3-Omni}}

\subsection{Evaluation library}
Our evaluation is fully decoupled from model serving: models run behind an HTTP
\emph{predict} endpoint, while the BEANS-Next runner performs post-processing and
metric computation locally. For each dataset example, the runner sends the audio
(and task prompt) to the model server, receives a raw text prediction, applies a
task-specific post-processing pipeline, and then computes deterministic metrics
(or routes to an LLM judge for non-deterministic tasks). The evaluation writes
both per-example artifacts (raw, processed, and scored predictions) and an
aggregate summary with per-task metrics.

GPT-4o audio-preview and Gemini 3.8 Flash were excluded due to high rates of API-level failures and cost, respectively. GPT-4o refused to process approximately 30\% of examples citing playback restrictions. For Gemini 3.8 Flash, we sampled 500 samples from all tasks in BEANS-Zero and BEANS-Next, and estimated the total cost for evaluating on the two benchmarks with `high' thinking mode to be \$7,500.

\subsection{Post-processing: parsing and label matching}
Let \( \hat{y}_{\text{raw}} \) be the raw model text output for an example.
Post-processing produces a \emph{segment list} \(S = [s_1,\dots,s_k]\) and a
canonical joined string \( \hat{y} \) obtained by concatenating segments with
``\texttt{, }''. The pipeline is explicitly split into two phases:

\paragraph{Parser phase (structural extraction).}
Parser steps only reshape text into label-like fragments without normalization.
For classification and detection tasks, we apply comma-based parsing:
\begin{itemize}
  \item \texttt{parse\_\allowbreak labels\_\allowbreak comma}: splits each segment on commas into
  multiple segments, while avoiding splitting numeric thousands separators
  (e.g., ``\texttt{1,654.75}'').
\end{itemize}

\paragraph{Cleaner phase (normalization and label alignment).}
Cleaner steps normalize segments and align them to the task label vocabulary:
\begin{itemize}
\raggedright
  \item \texttt{normalize\_\allowbreak whitespace}: collapses runs of whitespace and strips
  leading/trailing spaces per segment.
  \item \texttt{strip\_\allowbreak eos}: truncates any text following a model-specific
  end-of-sequence marker (default ``\texttt{<|end\_\allowbreak of\_\allowbreak text|>}'') per segment.
  \item \texttt{fuzzy\_\allowbreak match\_\allowbreak to\_\allowbreak labels}: maps each segment to the closest
  vocabulary label by minimum Levenshtein distance; optionally, a distance
  threshold can reject low-confidence matches (replacing them with a default
  label).
  \item \texttt{conservative\_\allowbreak match\_\allowbreak to\_\allowbreak labels}: a higher-precision variant
  intended for multi-label detection, using (i) case-insensitive exact match,
  (ii) case-insensitive longest substring match, then (iii) bounded Levenshtein
  matching; unmatched fragments are typically dropped.
  \item \texttt{extract\_\allowbreak label\_\allowbreak from\_\allowbreak text} / \texttt{extract\_\allowbreak mcq\_\allowbreak choice\_\allowbreak from\_\allowbreak text} /
  \texttt{extract\_\allowbreak scientific\_\allowbreak name\_\allowbreak from\_\allowbreak text} / \texttt{extract\_\allowbreak hz\_\allowbreak bucket\_\allowbreak from\_\allowbreak text}:
  specialized extractors used by tasks whose model outputs are verbose (e.g.,
  multiple-choice explanations, scientific-name outputs with timestamps, or
  numeric frequency estimates).
\end{itemize}

Fuzzy label matching is applied only when a task has a fixed label vocabulary
(e.g., classification/detection). For captioning and other free-form tasks
(open-ended, counting, QA), post-processing is limited to whitespace/EOS cleanup
to avoid corrupting natural-language outputs.

\subsection{Task types and metrics}
Each evaluation task declares a task type, which determines both the
post-processing configuration and the metric family.

\subsubsection{Classification}
Classification tasks produce a single label from a finite vocabulary. After
post-processing and label alignment, we compute:
\begin{itemize}
  \item \textbf{Accuracy}: fraction of examples where the (post-processed)
  predicted label exactly matches the ground-truth label.
  \item \textbf{Macro precision}, \textbf{macro recall}, \textbf{macro F1}:
  computed by averaging per-class precision/recall/F1 equally across classes.
\end{itemize}

\subsubsection{Detection (multi-label presence)}
Detection tasks allow multiple labels per example (presence/absence). After
comma parsing and label matching, we compute:
\begin{itemize}
  \item \textbf{Macro Average Precision (mAP)}: average precision computed per
  label from the precision--recall curve and then macro-averaged across labels.
  \item \textbf{Macro precision}, \textbf{macro recall}, \textbf{macro F1}:
  computed from predicted vs.\ target label sets (per example), then aggregated
  per task.
\end{itemize}

\subsubsection{Captioning}
Captioning tasks produce free-form text descriptions. Because single-sample CIDEr is degenerate, we compute captioning metrics at the corpus level:
\begin{itemize}
  \item \textbf{CIDEr (corpus mean, normalized to \([0,1]\))}: reported as the
  mean CIDEr score over the evaluation set.
\end{itemize}

\subsection{Computational resources}
For NatureLM-audio v1.0 and v1.1 checkpoints, as well as Audio Flamingo Next, an NVIDIA A100 40GB GPU was used for each for approximately 1 day to run all evaluations. For Qwen3-Omni, an NVIDIA H100 80GB GPU was used for approximately 1 day. These estimates are similar for the 3 audio-language models and 6 configurations of language models for the ablations in Sec. \ref{sec:text-only-baselines}. However, only \benchmark{} evaluation was done for these ablations, which accounts for around an eighth of samples. Thus, the total evaluation resources were around 4 A100 days and 1 H100 days for the main evaluations, and 1 A100 day and 0.5 H100 day for the ablations.

\FloatBarrier

\subsection{Text-only and Gaussian-noise controls}\label{sec:text-only-baselines}

MMAU and music question answering have indicated that LALMs can sometimes correctly respond to queries even when the correct audio is not provided~\citep{sakshi2024mmau, zang2025you}. Complementary experiments find that audio-language models can favor textual cues when text and audio conflict~\citep{wang-etal-2025-audio}. We thus evaluate the base LLMs of our chosen LALMs, namely Llama~3.1~8B, Qwen2.5~7B, and Qwen3~30B-A3B (although, Qwen3-Omni utilizes a modified version of it) without audio. We also measure performance of the chosen LALMs, namely NatureLM-audio v1.0, Audio Flamingo Next (AF-Next), and Qwen3-Omni, when Gaussian noise is provided as input, rather than the correct audio file. We use these experiments as proxies to assess whether textual cues are benefiting models during evaluation in \benchmark{}.

\paragraph{Conditions.}
We distinguish two text-only conditions. In the \emph{informed} condition, all audio inputs and placeholders are removed and the following instruction is prepended to each question:
\begin{quote}
\small
Audio is intentionally unavailable. Answer using only the text and answer choices. Make your best guess and follow the requested answer format. Do not refuse or explain.
\end{quote}
In the \emph{silent} condition, audio is removed without this instruction. The question text and answer choices are otherwise unchanged. We use each model's native chat template and deterministic decoding. For the three audio-language models, we compare real audio with duration-matched Gaussian noise at $-20$~dBFS RMS. Noise is supplied with the ordinary audio prompt, without an additional informed condition.

\paragraph{Scoring and answer validity.}
We apply the same task-specific scoring rules across conditions. Unambiguous answer prefixes and bulleted call-type lists are accepted; explicit missing-audio refusals are invalid answers rather than negative presence predictions. Invalid MCQ and binary-presence answers receive zero accuracy. Regression MAE is conditional on a parseable numerical answer, with its parse rate (the share of answers with a parseable number) reported alongside the error, as in Tables~\ref{tab:acoustic} and~\ref{tab:beansnext-t3-results}. For per-species counting, validity requires a parseable species--count mapping or explicit absence. Table~\ref{tab:ablation-fewshot-validity}(b) separately reports parse rates for MCQ and label-set tasks, pooling accepted answers over attempted examples.

\paragraph{Results.}
Tables~\ref{tab:ablation-tiers12}--\ref{tab:ablation-fewshot-validity} report the control results. Informing text models that audio is unavailable substantially changes their response behavior. Across Tier~1 MCQs, the three LLMs produce valid answers on 99.5--99.6\% of examples when informed, compared with 1.3--5.8\% in the silent condition. For Tier~3 MCQs, the corresponding ranges are 91.6--97.5\% and 0.3--1.6\%. Thus, an informed--silent performance difference reflects both answer validity and the usefulness of textual priors. Conversely, a valid answer can be uninformative: silent Qwen3 and Llama answer ``Yes'' on all 2,000 bird-presence examples, obtaining 50\% accuracy. Informed Qwen3 predicts all five call types on every example, yielding 46.4\% macro-F1 but zero exact-set accuracy. This baseline explains why a substantial multi-label score need not indicate sensitivity to the recording.

Real audio nevertheless provides a clear advantage on several tasks. Qwen3-Omni bird and amphibian presence accuracy falls from 90.0\% and 86.5\% with real audio to 50.0\% for both with noise; Audio Flamingo Next shows similar reductions, from 88.3\% and 80.4\% to 50.0\% and 49.5\%. Other comparisons are less decisive. Conditional regression errors can conceal low answer coverage: Audio Flamingo Next provides numerical estimates for only 1.8\% of held-out $f_0$ examples with real audio, while several no-audio conditions provide none. Tier~3 results also depend on output compatibility: only 0.05\% of NatureLM-audio v1.0 real-audio MCQ responses are parseable under the single-answer protocol. Its near-zero scores should therefore not be interpreted as a direct ranking of scene-understanding ability. Together, these controls show why task performance, answer validity, and simple constant-answer baselines should be considered jointly when assessing reliance on audio.

\providecommand{\prate}[1]{\,{\color{black!55}\scriptsize #1\%}}
\providecommand{\lowpr}[1]{{\color{black!55}#1}}
\begin{table}[p]
\centering
\caption{BEANS-Next acoustic and biological category recognition controls. Rows are grouped by input: real audio, duration-matched Gaussian noise, or no audio with (\emph{informed}) or without (\emph{silent}) the missing-audio instruction. Metrics and formatting follow Tables~\ref{tab:acoustic} and~\ref{tab:semantic}: higher is better unless marked $\downarrow$, small grey percentages give the parse rate, MAEs based on fewer than half of the answers are shown in grey, and -- marks no parseable answer. Insect and begging-call presence use 636 and 586 examples, respectively. $^{\dagger}$Retained inference failures: three Llama 3.1 8B seen-$f_0$ examples and one AF-Next caption per audio condition.}
\label{tab:ablation-tiers12}
\small
\renewcommand{\arraystretch}{1.1}
\textbf{(a) Tier 1: acoustic perception}

\smallskip
{\setlength{\tabcolsep}{3.2pt}
\begin{tabular}{@{}lrrrrrrr@{}}
\toprule
& \multicolumn{3}{c}{Acoustic description}
& \multicolumn{2}{c}{SNR}
& \multicolumn{2}{c}{Mean $f_0$ MAE (Hz) $\downarrow$} \\
\cmidrule(lr){2-4}
\cmidrule(lr){5-6}
\cmidrule(lr){7-8}
Model & Caption & Desc. MCQ & Crow & MCQ & MAE (dB) $\downarrow$ & Seen & Held-out \\
\midrule
\multicolumn{8}{@{}l}{\textit{Real audio}} \\
AF-Next & 0.4 & 50.1 & 47.5 & 33.9 & \lowpr{34.6}\prate{8.8} & \lowpr{5652}\prate{13.3} & \lowpr{249}\prate{1.8} \\
Qwen3-Omni & 0.0 & 47.0 & 45.5 & 7.7 & 20.0\prate{82.6} & 3728\prate{99.5} & 203\prate{99.1} \\
NatureLM-audio v1.0 & 0.3 & 23.6 & 33.5 & 31.9 & 24.9\prate{97.4} & 4087\prate{100} & 208\prate{100} \\
\addlinespace[3pt]
\multicolumn{8}{@{}l}{\textit{Gaussian noise}} \\
AF-Next & 0.8 & 25.6 & 36.5 & 30.4 & \lowpr{38.9}\prate{16.8} & \lowpr{4860}\prate{5.0} & \lowpr{--} \\
Qwen3-Omni & 0.0 & 13.5 & 25.0 & 4.3 & 42.4\prate{51.8} & 5440\prate{76.6} & \lowpr{254}\prate{21.7} \\
NatureLM-audio v1.0 & 0.5 & 21.8 & 25.5 & 28.5 & 36.1\prate{97.5} & 3773\prate{100} & 418\prate{100} \\
\midrule
\multicolumn{8}{@{}l}{\textit{No audio, informed}} \\
AF-Next & 0.0 & 43.2 & 38.0 & 32.5 & 24.3\prate{94.9} & 4492\prate{100} & 208\prate{100} \\
Qwen2.5 7B & 0.4 & 29.6 & 32.5 & 28.0 & 37.1\prate{90.5} & 4235\prate{100} & 79\prate{100} \\
Qwen3 30B-A3B & 0.5 & 22.9 & 32.5 & 22.4 & 16.6\prate{97.1} & \lowpr{--} & \lowpr{--} \\
Llama 3.1 8B & 0.1 & 27.5 & 34.5 & 38.1 & \lowpr{49.5}\prate{19.3} & \lowpr{--}$^{\dagger}$ & \lowpr{--} \\
\addlinespace[3pt]
\multicolumn{8}{@{}l}{\textit{No audio, silent}} \\
AF-Next & 0.0 & 40.2 & 37.0 & 32.2 & \lowpr{27.9}\prate{37.0} & \lowpr{--} & \lowpr{--} \\
Qwen2.5 7B & 0.0 & 0.1 & 11.5 & 0.0 & \lowpr{--} & \lowpr{--} & \lowpr{--} \\
Qwen3 30B-A3B & 0.0 & 0.0 & 30.5 & 0.0 & \lowpr{19.9}\prate{0.3} & \lowpr{--} & \lowpr{--} \\
Llama 3.1 8B & 0.0 & 0.0 & 8.5 & 0.0 & \lowpr{49.0}\prate{0.2} & \lowpr{--} & \lowpr{--} \\
\bottomrule
\end{tabular}}

\bigskip
\textbf{(b) Tier 2: biological category recognition}

\smallskip
{\setlength{\tabcolsep}{3pt}
\begin{tabular}{@{}lrrrrrrrrrr@{}}
\toprule
& \multicolumn{4}{c}{Taxon presence}
& \multicolumn{3}{c}{Call-type presence}
& & & \\
\cmidrule(lr){2-5}
\cmidrule(lr){6-8}
Model & Bird & Mamm. & Amph. & Insect & Alarm & Flight & Beg. & \shortstack{Call\\type} & Behav. & Caption \\
\midrule
\multicolumn{11}{@{}l}{\textit{Real audio}} \\
AF-Next & 88.3 & 49.5 & 80.4 & 75.8 & 59.6 & 55.6 & 54.4 & 44.1 & 44.6 & 0.0$^{\dagger}$ \\
Qwen3-Omni & 90.0 & 49.4 & 86.5 & 75.6 & 62.7 & 60.0 & 66.7 & 23.4 & 53.3 & 0.1 \\
NatureLM-audio v1.0 & 63.0 & 58.2 & 64.9 & 55.2 & 51.2 & 54.3 & 50.2 & 49.9 & 58.5 & 0.5 \\
\addlinespace[3pt]
\multicolumn{11}{@{}l}{\textit{Gaussian noise}} \\
AF-Next & 50.0 & 50.4 & 49.5 & 50.0 & 50.0 & 50.0 & 50.0 & 20.4 & 43.2 & 0.1$^{\dagger}$ \\
Qwen3-Omni & 50.0 & 50.0 & 50.0 & 50.2 & 50.0 & 50.0 & 50.0 & 13.7 & 48.7 & 0.1 \\
NatureLM-audio v1.0 & 51.1 & 50.0 & 50.9 & 50.2 & 50.0 & 50.1 & 50.0 & 15.3 & 46.2 & 0.0 \\
\midrule
\multicolumn{11}{@{}l}{\textit{No audio, informed}} \\
AF-Next & 50.0 & 50.0 & 50.0 & 50.0 & 50.0 & 50.0 & 50.0 & 46.4 & 34.0 & 0.0 \\
Qwen2.5 7B & 50.0 & 50.0 & 50.0 & 50.0 & 50.0 & 50.0 & 50.0 & 13.7 & 31.5 & 0.0 \\
Qwen3 30B-A3B & 50.0 & 50.0 & 50.0 & 50.0 & 50.0 & 50.0 & 50.0 & 46.4 & 33.2 & 0.0 \\
Llama 3.1 8B & 50.0 & 50.0 & 50.0 & 50.0 & 50.0 & 50.0 & 50.0 & 28.1 & 31.4 & 0.0 \\
\addlinespace[3pt]
\multicolumn{11}{@{}l}{\textit{No audio, silent}} \\
AF-Next & 50.0 & 50.0 & 50.0 & 50.0 & 50.0 & 50.0 & 50.0 & 46.4 & 39.2 & 0.1 \\
Qwen2.5 7B & 0.0 & 0.0 & 0.0 & 50.0 & 0.0 & 0.0 & 0.0 & 0.0 & 0.1 & 0.0 \\
Qwen3 30B-A3B & 50.0 & 50.0 & 50.0 & 50.0 & 50.0 & 50.0 & 50.0 & 0.0 & 3.2 & 0.0 \\
Llama 3.1 8B & 50.0 & 50.0 & 0.0 & 50.0 & 0.0 & 0.0 & 50.0 & 0.0 & 0.3 & 0.0 \\
\bottomrule
\end{tabular}}
\end{table}

\providecommand{\prate}[1]{\,{\color{black!55}\scriptsize #1\%}}
\providecommand{\lowpr}[1]{{\color{black!55}#1}}
\begin{table}[p]
\centering
\caption{BEANS-Next scene-understanding controls; inputs and formatting as in Table~\ref{tab:ablation-tiers12}, metrics as in Table~\ref{tab:beansnext-t3-results}. \texttt{\#V} MCQ (accuracy) and Presence (vocalization presence, accuracy) are additional Tier 3 tasks not shown in Table~\ref{tab:beansnext-t3-results}. For Range, unmatched or unparseable examples score zero. NatureLM-audio v1.0 segment-wise responses are scored as returned.}
\label{tab:ablation-tier3}
\small
\renewcommand{\arraystretch}{1.1}
\textbf{(a) Counting}

\smallskip
{\setlength{\tabcolsep}{6pt}
\begin{tabular}{@{}lrrrr@{}}
\toprule
& \multicolumn{3}{c}{Counting MAE $\downarrow$} & \\
\cmidrule(lr){2-4}
Model & \texttt{\#S} & \texttt{\#V} & \texttt{\#V}/\texttt{S} & \texttt{\#V} MCQ \\
\midrule
\multicolumn{5}{@{}l}{\textit{Real audio}} \\
AF-Next & 1.65\prate{97.4} & 4.23\prate{99.6} & 3.08\prate{53.2} & 26.0 \\
Qwen3-Omni & 1.20\prate{93.5} & 3.82\prate{99.8} & 1.92\prate{84.9} & 51.8 \\
NatureLM-audio v1.0 & \lowpr{1.00}\prate{41.3} & \lowpr{6.58}\prate{24.8} & \lowpr{2.11}\prate{1.7} & 0.0 \\
\addlinespace[3pt]
\multicolumn{5}{@{}l}{\textit{Gaussian noise}} \\
AF-Next & 2.38\prate{93.2} & 4.77\prate{100} & 6.08\prate{61.9} & 47.8 \\
Qwen3-Omni & 2.73\prate{81.5} & 5.24\prate{99.8} & \lowpr{1.29}\prate{10.9} & 54.2 \\
NatureLM-audio v1.0 & 1.11\prate{65.7} & 3.18\prate{83.4} & \lowpr{--} & 0.1 \\
\midrule
\multicolumn{5}{@{}l}{\textit{No audio, informed}} \\
AF-Next & 1.85\prate{100} & 7.59\prate{100} & \lowpr{3.31}\prate{33.9} & 15.4 \\
Qwen2.5 7B & 2.21\prate{100} & 4.89\prate{100} & \lowpr{5.37}\prate{33.9} & 33.5 \\
Qwen3 30B-A3B & 2.03\prate{100} & 4.68\prate{100} & \lowpr{2.69}\prate{2.9} & 29.0 \\
Llama 3.1 8B & \lowpr{1.94}\prate{43.1} & \lowpr{1.99}\prate{27.5} & \lowpr{2.17}\prate{32.6} & 20.4 \\
\addlinespace[3pt]
\multicolumn{5}{@{}l}{\textit{No audio, silent}} \\
AF-Next & 1.96\prate{100} & 5.40\prate{100} & \lowpr{--} & 13.6 \\
Qwen2.5 7B & \lowpr{2.07}\prate{30.4} & \lowpr{5.99}\prate{38.7} & \lowpr{1.94}\prate{32.1} & 0.6 \\
Qwen3 30B-A3B & \lowpr{--} & \lowpr{--} & \lowpr{2.04}\prate{33.9} & 0.0 \\
Llama 3.1 8B & 1.95\prate{51.1} & 4.06\prate{55.9} & \lowpr{2.37}\prate{49.2} & 0.0 \\
\bottomrule
\end{tabular}}

\bigskip
\textbf{(b) Temporal and scene-level tasks}

\smallskip
{\setlength{\tabcolsep}{6pt}
\begin{tabular}{@{}lrrrrrr@{}}
\toprule
& & \multicolumn{2}{c}{Temporal} & \multicolumn{3}{c}{Scene / open} \\
\cmidrule(lr){3-4}
\cmidrule(lr){5-7}
Model & Presence & Referring & Overlap & Range & Sum. & Caption \\
\midrule
\multicolumn{7}{@{}l}{\textit{Real audio}} \\
AF-Next & 50.9 & 33.5 & 47.3 & 0.0 & 0.0 & 0.0 \\
Qwen3-Omni & 51.2 & 25.0 & 48.9 & 0.0 & 0.0 & 0.0 \\
NatureLM-audio v1.0 & 49.7 & 0.0 & 58.4 & 0.0 & 0.1 & 0.1 \\
\addlinespace[3pt]
\multicolumn{7}{@{}l}{\textit{Gaussian noise}} \\
AF-Next & 50.0 & 18.9 & 55.2 & 0.0 & 0.0 & 0.0 \\
Qwen3-Omni & 50.0 & 21.4 & 54.9 & 0.0 & 0.0 & 0.1 \\
NatureLM-audio v1.0 & 61.5 & 0.2 & 54.9 & 0.0 & 0.0 & 0.0 \\
\midrule
\multicolumn{7}{@{}l}{\textit{No audio, informed}} \\
AF-Next & 50.0 & 22.5 & 55.4 & 0.0 & 0.0 & 0.0 \\
Qwen2.5 7B & 50.0 & 21.1 & 54.6 & 0.0 & 0.0 & 0.1 \\
Qwen3 30B-A3B & 50.0 & 25.6 & 51.3 & 0.0 & 0.1 & 0.0 \\
Llama 3.1 8B & 50.0 & 22.5 & 54.8 & 0.0 & 0.0 & 0.1 \\
\addlinespace[3pt]
\multicolumn{7}{@{}l}{\textit{No audio, silent}} \\
AF-Next & 50.0 & 24.9 & 54.4 & 0.0 & 0.0 & 0.0 \\
Qwen2.5 7B & 0.0 & 0.0 & 36.4 & 0.0 & 0.0 & 0.0 \\
Qwen3 30B-A3B & 31.0 & 0.3 & 14.7 & 0.0 & 0.0 & 0.0 \\
Llama 3.1 8B & 8.1 & 0.3 & 37.6 & 0.0 & 0.0 & 0.0 \\
\bottomrule
\end{tabular}}
\end{table}
\begin{table}[p]
\centering
\caption{BEANS-Next species-ID controls; inputs, metrics, and formatting as in Tables~\ref{tab:beansnext-t3-results} and~\ref{tab:ablation-tier3}. $^{\dagger}$Qwen2.5 7B silent lowest-pitch identification uses 799 of 803 responses.}
\label{tab:ablation-tier3-species}
\small
\renewcommand{\arraystretch}{1.1}
\textbf{Species ID by characteristic}

\smallskip
{\setlength{\tabcolsep}{4pt}
\begin{tabular}{@{}lrrrrrrrrrr@{}}
\toprule
& \multicolumn{5}{c}{Open-ended} & \multicolumn{5}{c}{MCQ} \\
\cmidrule(lr){2-6}
\cmidrule(lr){7-11}
Model & Order & High & Low & Long & Freq. & Order & High & Low & Long & Freq. \\
\midrule
\multicolumn{11}{@{}l}{\textit{Real audio}} \\
AF-Next & 0.0 & 0.0 & 0.0 & 0.0 & 0.0 & 27.0 & 32.2 & 46.0 & 32.6 & 24.5 \\
Qwen3-Omni & 0.0 & 0.0 & 0.0 & 0.0 & 0.0 & 25.7 & 33.6 & 19.7 & 19.2 & 13.4 \\
NatureLM-audio v1.0 & 0.0 & 0.0 & 0.0 & 0.0 & 0.0 & 0.1 & 0.0 & 0.0 & 0.0 & 0.0 \\
\addlinespace[3pt]
\multicolumn{11}{@{}l}{\textit{Gaussian noise}} \\
AF-Next & 0.0 & 0.0 & 0.0 & 0.0 & 0.0 & 26.8 & 31.7 & 40.1 & 30.2 & 23.0 \\
Qwen3-Omni & 0.0 & 0.0 & 0.0 & 0.0 & 0.0 & 27.6 & 24.0 & 18.2 & 14.0 & 10.5 \\
NatureLM-audio v1.0 & 0.1 & 0.1 & 0.0 & 0.1 & 0.2 & 2.2 & 3.8 & 4.7 & 3.8 & 2.2 \\
\midrule
\multicolumn{11}{@{}l}{\textit{No audio, informed}} \\
AF-Next & 0.0 & 0.0 & 0.0 & 0.0 & 0.0 & 22.4 & 33.4 & 43.1 & 29.4 & 21.0 \\
Qwen2.5 7B & 0.2 & 0.0 & 0.0 & 0.0 & 0.0 & 30.3 & 40.4 & 43.3 & 34.6 & 35.2 \\
Qwen3 30B-A3B & 0.0 & 0.0 & 0.0 & 0.0 & 0.0 & 31.9 & 46.7 & 45.1 & 38.7 & 20.4 \\
Llama 3.1 8B & 0.0 & 0.0 & 0.0 & 0.0 & 0.0 & 31.4 & 34.7 & 41.1 & 31.7 & 30.9 \\
\addlinespace[3pt]
\multicolumn{11}{@{}l}{\textit{No audio, silent}} \\
AF-Next & 0.0 & 0.0 & 0.0 & 0.0 & 0.0 & 24.2 & 32.6 & 39.7 & 26.5 & 22.0 \\
Qwen2.5 7B & 0.0 & 0.0 & 0.0$^{\dagger}$ & 0.0 & 0.0 & 0.5 & 0.0 & 0.0 & 0.0 & 0.1 \\
Qwen3 30B-A3B & 0.0 & 0.0 & 0.0 & 0.0 & 0.0 & 0.7 & 0.1 & 0.2 & 0.7 & 0.3 \\
Llama 3.1 8B & 0.0 & 0.0 & 0.0 & 0.0 & 0.0 & 0.2 & 0.0 & 0.2 & 0.0 & 0.0 \\
\bottomrule
\end{tabular}}
\end{table}

\providecommand{\prate}[1]{\,{\color{black!55}\scriptsize #1\%}}
\providecommand{\lowpr}[1]{{\color{black!55}#1}}
\begin{table}[p]
\centering
\caption{BEANS-Next in-context learning controls and parse rates; inputs and formatting as in Table~\ref{tab:ablation-tiers12}. (a) Gibbons and DCASE use macro-F1; other tasks use accuracy, all on a 0--1 scale. Zebra is an additional matching task not shown in Table~\ref{tab:multi-audio-fewshot}. NatureLM-audio v1.0 does not accept multiple audio inputs. (b) Parse rate (\%) for categorical answers, pooled over attempted examples: T1 MCQ ($n=3{,}468$, three tasks), T2 behavior ($3{,}657$), T3 MCQ ($5{,}807$, seven), T4 matching ($958$, four), call type ($2{,}000$), and T4 detection ($4{,}026$, two). A parseable choice or label set need not be correct; an explicit no-detection answer counts where allowed. Binary presence and captioning are excluded. --: not applicable.}
\label{tab:ablation-fewshot-validity}
\small
\renewcommand{\arraystretch}{1.1}
\textbf{(a) Tier 4: in-context learning}

\smallskip
{\setlength{\tabcolsep}{6pt}
\begin{tabular}{@{}lrrrrrr@{}}
\toprule
& \multicolumn{6}{c}{Few-shot task} \\
\cmidrule(lr){2-7}
Model & Gibbons & Otters & DCASE & Crows & Unseen & Zebra \\
\midrule
\multicolumn{7}{@{}l}{\textit{Real audio}} \\
AF-Next & 0.028 & 0.174 & 0.141 & 0.225 & 0.046 & 0.300 \\
Qwen3-Omni & 0.196 & 0.492 & 0.323 & 0.580 & 0.486 & 0.200 \\
\addlinespace[3pt]
\multicolumn{7}{@{}l}{\textit{Gaussian noise}} \\
AF-Next & 0.000 & 0.286 & 0.000 & 0.250 & 0.050 & 0.275 \\
Qwen3-Omni & 0.184 & 0.142 & 0.094 & 0.065 & 0.018 & 0.275 \\
\midrule
\multicolumn{7}{@{}l}{\textit{No audio, informed}} \\
AF-Next & 0.000 & 0.250 & 0.000 & 0.250 & 0.252 & 0.250 \\
Qwen2.5 7B & 0.199 & 0.250 & 0.061 & 0.250 & 0.252 & 0.250 \\
Qwen3 30B-A3B & 0.246 & 0.250 & 0.200 & 0.250 & 0.252 & 0.250 \\
Llama 3.1 8B & 0.000 & 0.000 & 0.000 & 0.000 & 0.000 & 0.000 \\
\addlinespace[3pt]
\multicolumn{7}{@{}l}{\textit{No audio, silent}} \\
AF-Next & 0.186 & 0.250 & 0.244 & 0.250 & 0.248 & 0.250 \\
Qwen2.5 7B & 0.000 & 0.000 & 0.000 & 0.000 & 0.000 & 0.000 \\
Qwen3 30B-A3B & 0.000 & 0.000 & 0.000 & 0.000 & 0.000 & 0.000 \\
Llama 3.1 8B & 0.000 & 0.000 & 0.000 & 0.000 & 0.000 & 0.000 \\
\bottomrule
\end{tabular}}

\bigskip
\textbf{(b) Parse rate for categorical answers}

\smallskip
{\setlength{\tabcolsep}{3pt}
\begin{tabular}{@{}lrrrrrr@{}}
\toprule
Model & T1 MCQ & T2 behavior & T3 MCQ & T4 matching & Call type & T4 detection \\
\midrule
\multicolumn{7}{@{}l}{\textit{Real audio}} \\
AF-Next & 98.1 & 99.9 & 91.3 & 69.1 & 67.5 & 95.6 \\
Qwen3-Omni & 54.4 & 97.9 & 55.5 & 87.7 & 100 & 81.6 \\
NatureLM-audio v1.0 & 97.4 & 92.9 & 0.1 & -- & 99.8 & -- \\
\addlinespace[3pt]
\multicolumn{7}{@{}l}{\textit{Gaussian noise}} \\
AF-Next & 99.1 & 99.8 & 84.2 & 82.8 & 100 & 48.9 \\
Qwen3-Omni & 80.9 & 99.8 & 54.1 & 48.7 & 100 & 24.0 \\
NatureLM-audio v1.0 & 96.9 & 99.8 & 3.7 & -- & 100 & -- \\
\midrule
\multicolumn{7}{@{}l}{\textit{No audio, informed}} \\
AF-Next & 99.2 & 99.8 & 85.1 & 100 & 100 & 100 \\
Qwen2.5 7B & 99.6 & 99.9 & 91.6 & 100 & 100 & 100 \\
Qwen3 30B-A3B & 99.6 & 99.9 & 97.5 & 100 & 100 & 100 \\
Llama 3.1 8B & 99.5 & 99.2 & 92.8 & 0.0 & 100 & 0.0 \\
\addlinespace[3pt]
\multicolumn{7}{@{}l}{\textit{No audio, silent}} \\
AF-Next & 98.8 & 99.8 & 84.8 & 100 & 100 & 100 \\
Qwen2.5 7B & 1.9 & 0.4 & 0.6 & 0.0 & 0.0 & 0.0 \\
Qwen3 30B-A3B & 5.8 & 8.2 & 1.6 & 0.0 & 0.0 & 0.0 \\
Llama 3.1 8B & 1.3 & 0.3 & 0.3 & 0.0 & 0.0 & 0.0 \\
\bottomrule
\end{tabular}}
\end{table}

\FloatBarrier

\section{Examples}
\begingroup
\footnotesize
\captionsetup{singlelinecheck=false}
\setlength{\tabcolsep}{4pt}
\renewcommand{\arraystretch}{1.08}
\setlength{\LTcapwidth}{\textwidth}
\setlength{\LTleft}{0pt}
\setlength{\LTright}{0pt}
\begin{longtable}{@{}>{\raggedright\arraybackslash}p{\dimexpr0.16\linewidth-1.5\tabcolsep\relax}>{\raggedright\arraybackslash}p{\dimexpr0.10\linewidth-1.5\tabcolsep\relax}>{\raggedright\arraybackslash}p{\dimexpr0.35\linewidth-1.5\tabcolsep\relax}>{\raggedright\arraybackslash}p{\dimexpr0.39\linewidth-1.5\tabcolsep\relax}@{}}
\caption{BEANS-Next benchmark examples. Audio placeholders are omitted or compacted; each row shows the split/task family, prompt, and reference output used for evaluation.}
\label{tab:beanspro_examples} \\
\toprule
\textbf{Split / task family} & \textbf{Format} & \textbf{Prompt} & \textbf{Reference output} \\
\midrule
\endfirsthead
\multicolumn{4}{l}{\tablename~\thetable{} (continued)} \\
\toprule
\textbf{Split / task family} & \textbf{Format} & \textbf{Prompt} & \textbf{Reference output} \\
\midrule
\endhead
\midrule
\multicolumn{4}{r}{\emph{Continued on next page}} \\
\endfoot
\bottomrule
\endlastfoot
T1 vocalization description & MCQ & Which description best matches the vocalization? Options include a slow trill, low drone, sharp whistle, and staccato notes with nasal rise. & a series of rapid, repetitive, staccato notes followed by a nasal, rising inflection \\
T1 vocalization description & MCQ & Which description best matches the sound? Options include impulsive bursts, a melodic whistle, sharp staccato notes, and a low nasal moan. & A series of dense, impulsive bursts followed by a quiet, rapid trill and punctuated by loud clicks \\
T1 acoustic caption & Caption & Describe the acoustic character of this vocalization, including its frequency range. & The vocalizations consist of high-pitched, whistling one-note sounds spanning roughly 2,500 to 7,000 Hz, as well as buzzing, nasal one-note sounds ranging from 2,000 to 18,000 Hz. \\
T1 acoustic caption & Caption & Describe the acoustic character of this vocalization, including its recording quality. & This vocalization consists of a series of rapid rattles that increase in frequency, concluding with a single prolonged rattle. The sound is difficult to hear over background noise but becomes more audible towards the end of the recording. \\
T2 behavior & MCQ & Which acoustic behavior is the Flame-crested Manakin performing? Options: wing snap, no vocalization/silent, song. & Wing snap \\
T2 behavior & MCQ & Which vocalizations can be heard from the blue duck? Options: rapid drumming, continuous song, male whistle followed by female growl. & A male whistle followed by a female growl \\
T2 semantic caption & Caption & What can you hear in this recording? & A pair of Bronze-winged Ducks produces distinct vocalizations. The female emits quacking calls, while the male provides accompanying wheezy whistles. \\
T2 semantic caption & Caption & Describe the vocalization in this recording. & A Scarlet-and-white Tanager pair performs a duet, consisting of a single note from the male and a fine rattle from the female. Faint background vocalizations from other species are also audible. \\
T2 semantic caption & Caption & What can be heard in this recording? & Multiple Tawny Owls engage in a complex vocal interaction consisting of compound hooting and ``ku-wick'' calls. The calls include high-pitched, harsh hooting from a female responding to at least three males. \\
T3 structural caption & Caption & Describe the pattern of vocalizations in this recording. & The recording features a sequence of low-frequency calls from \emph{Streptopelia decaocto}, which appear at the beginning, middle, and end of the clip. These calls are interspersed with higher-pitched vocalizations from \emph{Sturnus unicolor} and \emph{Luscinia megarhynchos}. \\
T3 highest pitch & MCQ & Which species produces the highest-pitched vocalization? Options: \emph{Charadrius hiaticula}, \emph{Numenius arquata}, \emph{Motacilla alba}, \emph{Erithacus rubecula}. & \emph{Motacilla alba} \\
T3 longest vocalization & MCQ & Which species produces the longest vocalization? Options: \emph{Poecile atricapillus}, \emph{Pipilo erythrophthalmus}, \emph{Cardinalis cardinalis}, \emph{Geothlypis trichas}. & \emph{Cardinalis cardinalis} \\
T3 co-occurrence & Binary & Are there any moments where both \emph{Vireo olivaceus} and \emph{Setophaga ruticilla} are calling simultaneously? & Yes \\
T3 species listing & Open-ended & How many different species are vocalizing, and what are they? Give scientific names. & 4: \emph{Corvus brachyrhynchos}, \emph{Pipilo erythrophthalmus}, \emph{Seiurus aurocapilla}, \emph{Setophaga virens} \\
T3 per-species counting & Open-ended & How many calls from each species can you hear? Give scientific names. & \emph{Pipilo erythrophthalmus}: 6; \emph{Cardinalis cardinalis}: 5; \emph{Geothlypis trichas}: 4; \emph{Polioptila caerulea}: 2 \\
T3 ordered summary & Open-ended & Describe each species in order of appearance, with call counts and frequency ranges. & \emph{Corvus brachyrhynchos}: 3 calls, 920--2130 Hz; \emph{Seiurus aurocapilla}: 2 calls, 2730--9550 Hz; \emph{Regulus calendula}: 2 calls, 1430--8030 Hz; \emph{Poecile atricapillus}: 6 calls, 2850--8830 Hz. \\
T4 gibbon few-shot detection & Multi-audio & Given three support sounds A--C, which are present in the query recording, if any? & Multiple pulse gibbon call \\
T4 giant otter call matching & Multi-audio & Given four call-type audio exemplars A--D, which call type best matches the query recording? & Begging Scream \\
T4 held-out species matching & Multi-audio & Given four species audio exemplars A--D, which species best matches the query recording? & \emph{Minervarya agricola} \\
\end{longtable}
\endgroup

\clearpage

\end{document}